\documentclass[traditabstract,longauth]{aa}

\usepackage{graphicx}

\usepackage{txfonts}

\usepackage{natbib}
\bibpunct{(}{)}{;}{a}{}{,} % to follow the A&A style
\defcitealias{leger2009}{L{\'e}ger, Rouan, Schneider et al. 2009}

\newcommand{\corot}{\textsl{CoRoT}}

\def\ms{\hbox{\,m\,s$^{-1}$}}         %m.s-1
\def\m2s2{\hbox{\,m$^{2}$\,s$^{-2}$}} %m2.s-2
\def\kms{\hbox{\,km\,s$^{-1}$}}       %km.s-1
\def\gcm3{\hbox{\,g\,cm$^{-3}$}}      %g.cm-3
\def\vsini{\hbox{$v$\,sin\,$i$}}      %vsini
\def\Msun{\hbox{$M_{\odot}$}}         %Msun
\def\Rsun{\hbox{$R_{\odot}$}}
\def\Mjup{\hbox{$\mathrm{M}_{\rm Jup}$}}
\def\Rjup{\hbox{$\mathrm{R}_{\rm Jup}$}}
\def\Mearth{\hbox{$\mathrm{M}_{\oplus}$}}

\begin{document}

% http://www.eso.org/sci/observing/policies/publications.html
\title{Transiting exoplanets from the CoRoT space mission
  \thanks{The CoRoT space mission, launched on December 27th 2006, has
  been developed and is operated by CNES, with the contribution of
  Austria, Belgium, Brazil , ESA (RSSD and Science Programme), Germany
  and Spain. 
% CFHT
  Part of the observations were obtained at the Canada-France-Hawaii
  Telescope (CFHT) which is operated by the National Research Council
  of Canada, the Institut National des Sciences de l'Univers of the
  Centre National de la Recherche Scientifique of France, and the
  University of Hawaii.
% HARPS@ESO
%  Based on observations collected at the European
%  Organisation for Astronomical Research in the Southern Hemisphere,
%  Chile (ESO program 184.C-0639). 
  Based on observations made with HARPS spectrograph on the
  3.6-m European Organisation for Astronomical Research in the
  Southern Hemisphere telescope at La Silla Observatory, Chile (ESO
  program 184.C-0639).
% IAC80
  Based on observations made with the IAC80 telescope
  operated on the island of Tenerife by the Instituto de
  Astrof{\'i}sica de Canarias in the Spanish Observatorio del Teide.
% Keck
  Part of the data presented herein were obtained at the W.M. Keck
  Observatory, which is operated as a scientific partnership among the
  California Institute of Technology, the University of California and
  the National Aeronautics and Space Administration. The Observatory
  was made possible by the generous financial support of the W.M. Keck
  Foundation. 
  }
}

%\subtitle{XVI. CoRoT-13b: the lucky planet}
%\subtitle{XVI. CoRoT-13b: a hot Jupiter in transit around a star with super-solar lithium content}
\subtitle{XIII. CoRoT-13b: a dense hot Jupiter in transit around a star
  with solar metallicity and super-solar lithium content} 

\author{
J.~Cabrera\inst{\ref{DLR},\ref{LUTh}} 
\and H.~Bruntt\inst{\ref{LESIA}}
\and M.~Ollivier\inst{\ref{IAS}} 
\and R.~F.~D{\'i}az\inst{\ref{IAP}}
\and Sz.~Csizmadia\inst{\ref{DLR}} 
%%%% 
\and S.~Aigrain\inst{\ref{Oxford}} 
\and R.~Alonso\inst{\ref{Geneve}} 
\and J.-M.~Almenara\inst{\ref{IAC}} 
\and M.~Auvergne\inst{\ref{LESIA}} 
\and A.~Baglin\inst{\ref{LESIA}}
\and P.~Barge\inst{\ref{LAM}} 
\and A.~S.~Bonomo\inst{\ref{LAM}}  
\and P.~Bord\'e\inst{\ref{IAS}} 
\and F.~Bouchy\inst{\ref{OHP},\ref{IAP}} 
\and L.~Carone\inst{\ref{Koeln}} 
\and S.~Carpano\inst{\ref{ESA}} 
\and M.~Deleuil\inst{\ref{LAM}} 
\and H.~J.~Deeg\inst{\ref{IAC}} 
\and R.~Dvorak\inst{\ref{Wien}} 
\and A.~Erikson\inst{\ref{DLR}}
\and S.~Ferraz-Mello\inst{\ref{Brasil}} 
\and M.~Fridlund\inst{\ref{ESA}}
\and D.~Gandolfi\inst{\ref{Tautenburg},\ref{ESA}}
\and J.-C.~Gazzano\inst{\ref{LAM},\ref{OCA}}
\and M.~Gillon\inst{\ref{Geneve},\ref{Liege}} 
\and E.~W.~Guenther\inst{\ref{Tautenburg}} 
\and T.~Guillot\inst{\ref{OCA}} 
\and A.~Hatzes\inst{\ref{Tautenburg}} 
\and M.~Havel\inst{\ref{OCA}}
\and G.~H\'ebrard\inst{\ref{IAP}} 
\and L.~Jorda\inst{\ref{LAM}} 
\and A.~L\'eger\inst{\ref{IAS}} 
\and A.~Llebaria\inst{\ref{LAM}} 
\and H.~Lammer\inst{\ref{Graz}} 
\and C.~Lovis\inst{\ref{Geneve}} 
\and T.~Mazeh\inst{\ref{Tel Aviv}} % \inst{\ref{Wise}} ?
\and C.~Moutou\inst{\ref{LAM}} 
\and A.~Ofir\inst{\ref{Tel Aviv}}
\and P.~von Paris\inst{\ref{DLR}}
\and M.~P\"atzold\inst{\ref{Koeln}} 
\and D.~Queloz\inst{\ref{Geneve}}
\and H.~Rauer\inst{\ref{DLR},\ref{ZAA}} 
\and D.~Rouan\inst{\ref{LESIA}}
\and A.~Santerne\inst{\ref{LAM}} 
\and J.~Schneider\inst{\ref{LUTh}} 
\and B.~Tingley\inst{\ref{IAC}} 
\and R.~Titz-Weider\inst{\ref{DLR}}
\and G.~Wuchterl\inst{\ref{Tautenburg}} 
}

\institute{
Institute of Planetary Research, German Aerospace Center, Rutherfordstrasse 2, 12489 Berlin, Germany\label{DLR}
\and LUTH, Observatoire de Paris, UMR 8102 CNRS, Universit\'e Paris Diderot; 5 place Jules Janssen, 92195 Meudon, France\label{LUTh}
\and LESIA, Observatoire de Paris, Place Jules Janssen, 92195 Meudon cedex, France\label{LESIA}
\and Institut d'Astrophysique Spatiale, Universit\'e Paris XI, F-91405 Orsay, France\label{IAS}
\and Institut d'Astrophysique de Paris, UMR 7095 CNRS, Universit\'e Pierre \& Marie Curie, 98bis boulevard Arago, 75014 Paris, France"\label{IAP}
%\and Institut d'Astrophysique de Paris, 98bis boulevard Arago, 75014 Paris, France\label{IAP}
\and Department of Physics, Denys Wilkinson Building Keble Road, Oxford, OX1 3RH\label{Oxford}
\and Observatoire de l'Universit\'e de Gen\`eve, 51 chemin des Maillettes, 1290 Sauverny, Switzerland\label{Geneve}
\and Instituto de Astrof{\'i}sica de Canarias, E-38205 La Laguna, Tenerife, Spain\label{IAC}
\and Laboratoire d'Astrophysique de Marseille, 38 rue Fr\'ed\'eric Joliot-Curie, 13388 Marseille cedex 13, France\label{LAM}
\and Observatoire de Haute Provence, 04670 Saint Michel l'Observatoire, France\label{OHP}
\and Rheinisches Institut f\"ur Umweltforschung an der Universit\"at zu K\"oln, Aachener Strasse 209, 50931, Germany\label{Koeln}
\and Research and Scientific Support Department, ESTEC/ESA, PO Box 299, 2200 AG Noordwijk, The Netherlands\label{ESA} 
\and University of Vienna, Institute of Astronomy, T\"urkenschanzstr. 17, A-1180 Vienna, Austria\label{Wien}
\and IAG-Universidade de Sao Paulo, Brasil\label{Brasil}
%\and Observat\'orio Nacional, Rio de Janeiro, RJ, Brazil\label{Brasil}
\and Th\"uringer Landessternwarte, Sternwarte 5, Tautenburg 5, D-07778 Tautenburg, Germany\label{Tautenburg}
\and Universit\'e de Nice-Sophia Antipolis, CNRS UMR 6202, Observatoire de la C\^ote d'Azur, BP 4229, 06304 Nice Cedex 4, France\label{OCA}
%\and Observatoire de la C\^ote d'Azur, Laboratoire Cassiop\'ee, BP 4229, 06304 Nice Cedex 4, France\label{OCA}
\and University of Li\`ege, All\'ee du 6 ao\^ut 17, Sart Tilman, Li\`ege 1, Belgium\label{Liege}
\and Space Research Institute, Austrian Academy of Science, Schmiedlstr. 6, A-8042 Graz, Austria\label{Graz}
\and School of Physics and Astronomy, Raymond and Beverly Sackler Faculty of Exact Sciences, Tel Aviv University, Tel Aviv, Israel\label{Tel Aviv}
\and Center for Astronomy and Astrophysics, TU Berlin, Hardenbergstr. 36, 10623 Berlin, Germany\label{ZAA}
%\and School of Physics, University of Exeter, Stocker Road, Exeter EX4 4QL, United Kingdom\label{Exeter}
%\and Wise Observatory, Tel Aviv University, Tel Aviv 69978, Israel\label{Wise}
%\and Dpto. de Astrof\'isica, Universidad de La Laguna, 38206 La Laguna, Tenerife, Spain\label{La Laguna}
%\and Laboratoire d'Astronomie de Lille, Universit\'e de Lille 1, 1 impasse de l'Observatoire, 59000 Lille, France\label{Lille}
%\and Institut de M\'ecanique C\'eleste et de Calcul des Eph\'em\'erides, UMR 8028 du CNRS, 77 avenue Denfert-Rochereau, 75014 Paris, France\label{IMCCE}
}
\date{Received ; accepted }

% traditional abstract format
%% %% \abstract{We announce the discovery of the transiting planet
%% %%   CoRoT-13b. A hot Jupiter-like planet with an orbital period of
%% %%   $4.035\,190 \pm 0.000\,030$ days and a mass of $1.308 \pm 0.066$
%% %%   Jupiter masses. It orbits a G0V star with a metallicity of $+0.01
%% %%   \pm 0.07$ and a lithium content of $+1.45$ dex.
%% \abstract{We announce the discovery of the transiting planet
%%   CoRoT-13b. A hot Jupiter-like planet with an orbital period of
%%   $4.04$ days, $1.3$ Jupiter masses, and a density of
%%   $2.34\;g\;cm^{-3}$. It orbits a G0V star with a metallicity of 
%%   $+0.01 \pm 0.07$ and a lithium content of $+1.45$ dex.
%% \keywords{stars: planetary systems - techniques: photometry - techniques:
%%   radial velocities - techniques: spectroscopic }
%% }
%% 
\abstract{We announce the discovery of the transiting planet 
 CoRoT-13b. 
 Ground based follow-up in CFHT and IAC80 confirmed CoRoT's
 observations.
 The mass of the planet was measured with the HARPS spectrograph and
 the properties of the host star were obtained analyzing HIRES spectra
 from the Keck telescope.
 It is a hot Jupiter-like planet with an orbital period of $4.04$
 days, $1.3$ Jupiter masses, $0.9$ Jupiter radii, and a density of 
 $2.34$\gcm3.  
 It orbits a G0V star with $T_\mathrm{eff}=5\,945$K,
 $M_{*}=1.09$\Msun, $R_{*}=1.01$\Rsun, solar metallicity, a lithium
 content of $+1.45$ dex, and an estimated age between $0.12$ and
 $3.15$ Gyr.
 The lithium abundance of the star is consistent with its effective
 temperature, activity level, and age range derived from the stellar
 analysis. 
 The density of the planet is extreme for its mass. It implies the
 existence of an amount of heavy elements with a mass between about
 $140$ and $300$\Mearth.

}

%% % new abstract format
%% \abstract
%% % context heading (optional)
%% % {} leave it empty if necessary  
%% {}
%% % aims heading (mandatory)
%% {We announce the discovery of the transiting planet CoRoT-13b.}
%% % methods heading (mandatory)
%% {The transiting planet was discovered with the satellite CoRoT. It was
%%   confirmed with ground based photometric follow-up in CFHT and 
%%   IAC80. Its mass was measured with HARPS observations. The properties
%%   of the host star were obtained analyzing Keck data.} 
%% % results heading (mandatory)
%% {CoRoT-13b is a hot Jupiter-like planet with an orbital period of
%%   $4.04$ days, $1.3$ Jupiter masses, and a density of
%%   $2.34$\gcm3. It orbits a G0V star with a metallicity of
%%   $+0.01 \pm 0.07$ and a lithium content of $+1.45$ dex} 
%% % conclusions heading (optional), leave it empty if necessary 
%% {}
%% \keywords{stars: planetary systems - techniques: photometry - techniques:
%%    radial velocities - techniques: spectroscopic }

\titlerunning{}
\authorrunning{}

\maketitle

%
%____________________________________________________________________________
\section{Introduction}
\label{sec:introduction}

Transiting planets are fundamental objects for the understanding of
planetary formation and evolution. 
Their particular geometrical orientation allows the simultaneous
measurement of their mass and radius, permitting a first order study
of their internal structure. 
Moreover, through the careful analysis of their passages in front of
(primary transit) and behind (occultation of the planet or secondary
transit) their host star, a characterization of the composition and
the temperature structure of their atmosphere can be carried
out.%performed.

\corot\, is a space telescope dedicated to the study of
asteroseismology and the discovery of extrasolar planets by the method
of transits \citep{baglin2006}. 
%\corot\, survey stands out by its capability to 
%detect the faint signal of small rocky planets
%\citep{leger2009,queloz2009} and giant planets with temperate surface
%temperature \citep{deeg2010}.
The \corot\,survey has previously detected, among other planets, the
faint signal of a small rocky planet
(CoRoT-7b; \citetalias{leger2009}; \citealt{queloz2009,bruntt2010b})
and a giant planet with a temperate surface temperature
\citep[CoRoT-9b;][]{deeg2010}.

This paper presents the discovery of the planet CoRoT-13b, a dense
giant planet in a close orbit around a main sequence star. 
Section~\ref{sec:corot_lc} describes the observations performed with
the satellite. 
Ground based observations including spectroscopic characterization of
the star and radial velocity measurements are reported in
section~\ref{sec:ground_observations}. 
The derivation of the planetary parameters is described in
section~\ref{sec:planetary_parameters}. 
A discussion of the results is presented in section
\ref{sec:discussion}.

%
%____________________________________________________________________________
\section{CoRoT observations}
\label{sec:corot_lc}

The observations of the \corot\,field LRa02 started on November 16th
2008; the first transits of CoRoT-13b were discovered by the Alarm
Mode and the target was oversampled (the standard 512s sampling rate
was changed to 32s; see \citealt{surace2008}) since December 9th 2008
and until the end of the observations in March 11th 2009, gathering in
total about $250\,000$ photometric measurements.  
% Claire suggest to remove this sentence
%% CoRoT-13b turned
%% out to be a hot-Jupiter planet with an orbital period of $4.035\,190
%% \pm 0.000\,030$ days and a mass of $1.308 \pm 0.066$ Jupiter
%% masses. The final parameters of the system are reported in
%% Table~\ref{starplanet_param_table}.  

\begin{figure*}[t]
  \begin{center}
    \includegraphics[%
%      draft,%
      width=0.9\linewidth,%
      height=0.5\textheight,%
%      viewport=50 50 410 302,%
      keepaspectratio]{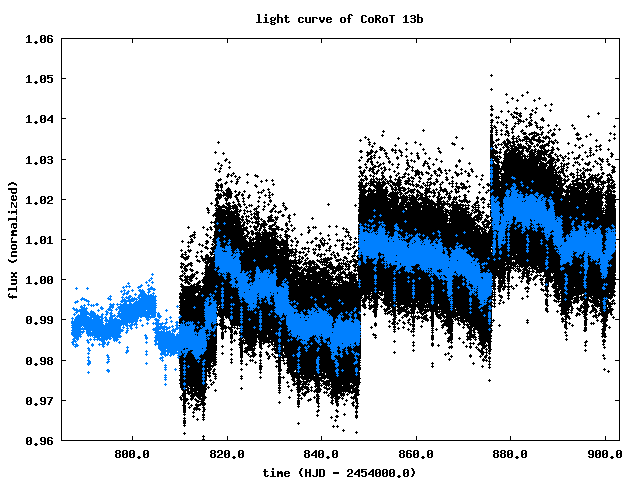}
  \end{center}
  \caption{
    Raw light curve of CoRoT-13b. The target was oversampled since
    December 9th 2008 (HJD $2\,454\,810.20$); for visualization
    purposes we superimpose the LC binned at a sampling rate of 512s
    to the 32s sampled region. The jumps in the data are from hot
    pixels as described in the text.
  }
  \label{fig:full_lc}
\end{figure*}

% table IDs, coordinates and magnitudes
\begin{table}
\caption{IDs, coordinates and magnitudes of CoRoT-13.}
\centering
\begin{tabular}{lcc}       
\hline\hline                 
CoRoT window ID & LRa02\_E2\_2165  \\
CoRoT ID        & 110839339        \\
UCAC3           & 170-048045       \\
USNO-A2 ID      & 0825-03324928    \\ 
USNO-B1 ID      & 0849-0108714     \\ 
2MASS ID        & 06505307-0505112 \\
\\
\multicolumn{2}{l}{Coordinates} \\
\hline            
RA (J2000)  &  6h 50m 53.07s         \\
Dec (J2000) &  -5$^\circ$ 5' 11.26'' \\
\\
\multicolumn{3}{l}{Magnitudes} \\
\hline
\centering
Filter & Mag & Error \\
\hline
B\tablefootmark{a}  & 15.777 & 0.077 \\
V\tablefootmark{a}  & 15.039 & 0.041 \\
r'\tablefootmark{a} & 14.738 & 0.027 \\
i'\tablefootmark{a} & 14.304 & 0.033 \\
J\tablefootmark{b}  & 13.710 & 0.021 \\
H\tablefootmark{b}  & 13.406 & 0.027 \\
K\tablefootmark{b}  & 13.376 & 0.038 \\
%\\                                    
%\multicolumn{3}{l}{Proper motion} \\
%\hline
%$\mu_{\alpha}$ &   8.0 & ''/yr
%$\mu_{\delta}$ & -11.8 & ''/yr
\hline
\end{tabular}
\tablefoot{
  \tablefoottext{a}{Provided by ExoDat \citep{deleuil2009};}
  \tablefoottext{b}{from 2MASS catalog.}
}
\label{startable}      
\end{table}

The coordinates, identification labels and magnitudes of CoRoT-13
are given in Table~\ref{startable}. 
It is a relatively faint object for \corot\,($V=15.039$) and it was
therefore assigned a monochromatic aperture mask (for a complete
description of the observing modes of the satellite, please refer to 
\citealt{boisnard2006,barge2008b,auvergne2009}).
%Figure~\ref{fig:fov} shows the area of the sky around the target: there are several
%contaminants whose light is mixed with that of the target in 
%\corot\,data. The contribution of the main contaminants has been
%substracted when modeling the transit (see discussion below).
%%, but its flux cannot be deblended in the \corot\,light curve (LC).
Figure~\ref{fig:full_lc} shows the raw light curve (LC) of the
target. 
One of the most significant environmental effects in \corot\,LCs are
hot pixels: proton impacts that produce permanent damage to the CCD
lattice (see \citealt{drummond2008,pinheirodasilva2008}). On average,
0.3\% of the pixels are affected by one of these events with an impact
of more than 1000 electrons in flux during a long run
\citep{auvergne2009}. 
The mask of CoRoT-13b has only 69 pixels, but unfortunately this
particular LC shows at least 3 impacts over the 1000 electrons
threshold plus some other impacts of less importance.
% In particular, the event around the \corot\,Julian date 3303.02 is
% present in all the LC in the neighborhood (up to 1 arc min) of the
% target. It is also visible in the background correction, which
% indicates that this is probably an impact in the region used for the
% background calculation.
These impacts do not affect significantly the characterization of
the target as long as the regions affected are properly treated 
\citep{pinheirodasilva2008}.
%but the overall highly noisy environment of this particular candidate
%affects the study of the planetary system scenario.

%____________________________________________________________________________
\section{Ground-based observations}
\label{sec:ground_observations}

%............................................................................
\subsection{Photometric measurements}
\label{subsec:photometric_measurements}

\begin{figure}[t]
  \begin{center}
    \includegraphics[%
      width=0.7\linewidth,%
      height=0.5\textheight,%
%      keepaspectratio]{fov}
      angle=270,
%      keepaspectratio]{Fig_Corot13_overlay_lite}
      keepaspectratio]{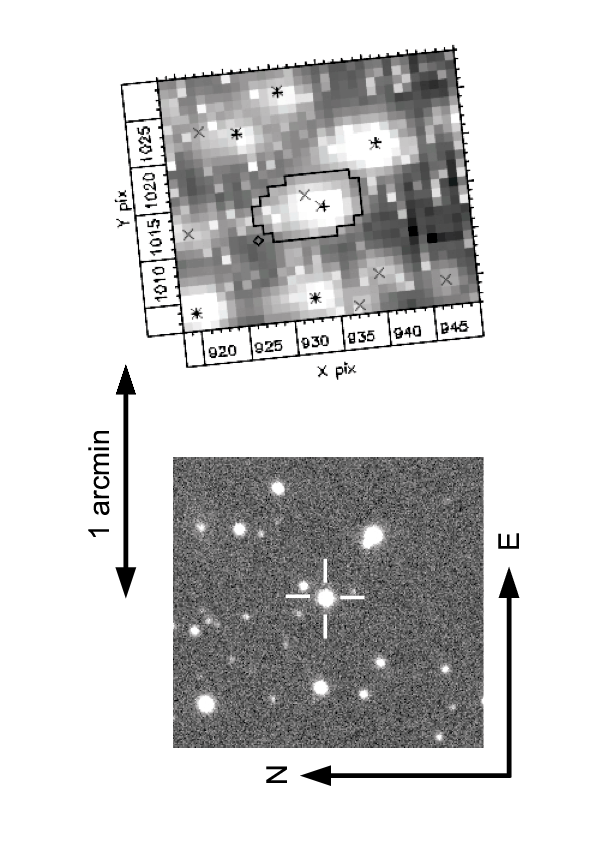}
  \end{center}
  \caption{
    Region of the sky around CoRoT-13. {\em Left:} Image taken with 
%    INT/WFC with a resolution of 1.3''. The position of the target is %%%%%%% ?????
    INT/WFC with a resolution of 0.3''. The position of the target is 
    indicated with a white cross. {\em Right:} Image taken by CoRoT
    with the same scale and orientation; the resolution is around
    2.32''/pixel. The area inside the black line is the mask used to
    compute the photometry; the position of the nearest contaminants
    from the ExoDat database are indicated with different types of
    crosses.
  }
  \label{fig:fov}
\end{figure}

A complete description of the photometric follow-up of \corot\,targets
is given in \citet{deeg2009}. 
For the particular case of CoRoT-13b, photometric measurements of the
host star were carried out at the 3.6m Canada-France-Hawaii-Telescope
(CFHT) in late November 2009 confirming that the transit was
on-target, although with a depth slightly bigger than expected. 
This is due to the contribution of the contaminants to the flux in the
\corot\, mask. 
Later on, in early January 2010, new on-off observations were carried
out at the IAC80 telescope, in Tenerife, confirming the previous
results.
% email D. Rouan 23Nov09: Last night CFHT observation of
% LRa02\_E2\_2165 gave the following result : The only star on which a
% flux variation with the correct sign is seen would be the target
% star, however, a) the dip is at least two times larger than expected
% (2.2 \% +/- 0.4 rather than .83 \%) and b) there is a strange
% bimodal distribution of the flux on both the ON and the OFF
% periods. I don't think that flares with a few minutes time scale are
% credible. If one separates the high state from the low state, the
% difference ON-OFF is however about the same : 1.5\% and 2.0
% \%. Given the separation (20 arcsec) of the brightest neighbour star
% (0.4 mag brighter), I don't think that dilution could be the
% explanation of the difference with the Corot LC. 

% Mail by HD to DR on 11 dec09: Since the on-target detection is
% somewhat unclear, i checked on nearby stars, and there are two
% around the faint limit (~19.3mag, measured from WFC images). These
% could in principle be false-alarm source if they vary very strongly
% >~0.5mag; see red circles in attached img. Could you please check on
% these, so that we can exclude that possiblity. 

% 22Jan10: DR sends to HD two lightcurves of the faint stars that show
% absence of relevent on-off variatations -> ON-target 

% 27Jan10: IAC80 on-off photometry (13jan10/14jan10) confirms an event
% with a depth slightly over 1\% in the target star, with no evidence
% of anything of interest in nearby stars (BT). 

Figure~\ref{fig:fov} shows the region of the sky around the target
taken with the Wide Field Camera of the Isaac Newton Telescope
(INT/WFC), in Roque de los Muchachos. 
The closest contaminants to the main target, according to the ExoDat
database \citep{deleuil2009}, are shown in
Table~\ref{table:contaminants}. 
In particular, the object 110839426 is completely included in the
\corot\,mask, diluting the transit signal of CoRoT-13b. 
An analysis of the point spread function (PSF) of \corot\,reveals that
the main target, the planet-hosting star, is responsible for
$89\pm1\%$ of the flux within the mask.
%; fact which has been taken into account for the modeling.

\begin{table}
  \caption{Closest contaminants to CoRoT-13b with their respective
    magnitudes and relative distances to the target.}
  \centering
  \begin{tabular}{*{5}{c}}
    \hline\hline
    CoRoT ID  & distance  & B        & V        & R        \\
              & (arc sec) &          &          &          \\
    \hline
    110839426 &  $6.58$   & $19.354$ & $18.136$ & $17.625$ \\
    110839769 & $20.14$   & $15.424$ & $14.689$ & $14.395$ \\
    110838938 & $21.58$   & $19.691$ & $18.412$ & $17.838$ \\
    110838780 & $22.79$   & $16.927$ & $16.066$ & $15.701$ \\
    110838726 & $28.34$   & $19.911$ & $18.489$ & $17.875$ \\
    110839832 & $28.44$   & $17.611$ & $16.749$ & $16.388$ \\
    \hline
  \end{tabular}
  \label{table:contaminants}      
\end{table}

%............................................................................
\subsection{Spectroscopic measurements}
\label{subsec:spectroscopic_measurements}

\begin{figure*}[t]
  \begin{center}
    \includegraphics[%
      angle=90,%
      width=0.9\linewidth,%
      height=0.5\textheight,%
%      bb=42 42 382 807, clip,%
      viewport=42 42 382 807,%
      keepaspectratio]{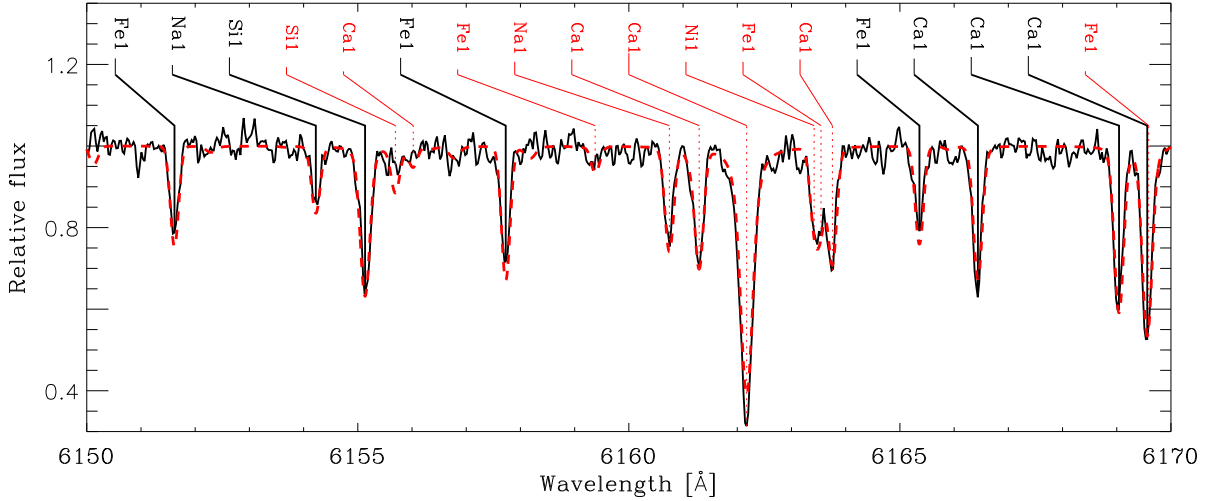}
  \end{center}
  \caption{
    Detail of the Keck spectrum of CoRoT-13 around the
    pressure-sensitive Ca {\sc ii} line at 6162 Å. The synthetic
    spectrum is shown with a dashed line. Spectral lines used in the
    abundance analysis are marked with solid vertical lines while the
    other spectral lines are marked with dotted lines.
  }
  \label{fig:spec}
\end{figure*}

\begin{figure}[t]
  \begin{center}
    \includegraphics[%
      angle=90,%
      width=0.9\linewidth,%
      height=0.5\textheight,%
%      bb=21 113 205 679, clip,%
%      viewport=21 113 205 679,%
      viewport=21 113 261 679,%
      keepaspectratio]{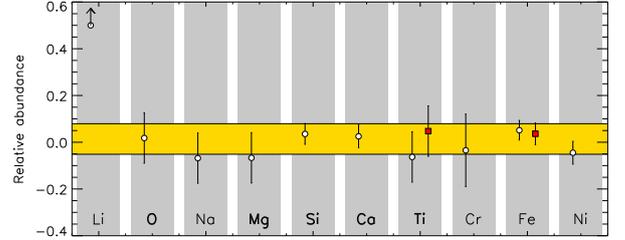}
  \end{center}
  \caption{
    Mean abundances for 10 elements in CoRoT-13 Keck
    spectrum. White circles correspond to neutral lines and red boxes
    stand for singly ionized lines. The yellow area represents the
    mean metallicity within one sigma error bar. 
  }
  \label{fig:abund}
\end{figure}

The star was observed with HIRES at the Keck on Dec. 5th 2009 as part
of the NASA's key science programme in support of the \corot\,mission.
% Observations were made for 1200 sec without the iodine cell. 
We obtained one spectrum without the iodine cell and with an exposure
time of 1200 sec. 
The spectral resolution is $\sim 45\,000$. 
To determine the atmospheric parameters of CoRoT-13 we analyzed the
spectrum using the VWA software \citep{bruntt2004, bruntt2010a}. 
We selected lines in the range 5050--7810\,\AA. 
A small section of the spectrum is shown in Fig.~\ref{fig:spec} where
the typical signal-to-noise (SN) ratio is 55. 
To determine the atmospheric parameters we use the neutral and ionized
Fe lines and also the wide Ca lines at 6122 and 6162\,\AA\ (see
\citealt{bruntt2010a} for a description). 
From this analysis we determine the parameters 
$T_{\rm eff}  =  5945 \pm 90$\,K,  
$\log g  =  4.30 \pm 0.10$, 
${\rm [M/H]} = +0.01 \pm 0.07$ (mean of Si, Fe and Ni), and projected 
rotational velocity $v \sin i = 4 \pm 1$\,km/s. 
The $T_{\rm eff}$ has been adjusted by $-40$~K based on the comparison
of 10 stars with $T_{\rm eff}$ determined from both interferometric
and spectroscopic methods as described by \cite{bruntt2010a}. 
The abundances of 10 elements are given in Table~\ref{table:abund} and
shown in Fig.~\ref{fig:abund}. 
The horizontal yellow bar is the mean metallicity. 
% added by Malcolm
Although the metallicity of this star is essentially solar, it is
notable that the abundance of {Li \sc i} is $+1.45$ dex, in similarity
to the case of another planet hosting star, CoRoT-6
\citep{fridlund2010}.

The mass, radius and age of the star are calculated using the
Starevol evolutionary tracks 
\citep[][Palacios, private communication]{siess2006}. The input
values are the spectroscopic parameters derived above (T$_{\rm eff}$  
and [Fe/H]) and the proxy for the stellar density
($M_\star^{1/3}/R_\star$) obtained from the modeling of the transit
light-curve. The results are 
$M_{*} = 1.09 \pm 0.02$\Msun; 
$R_{*} = 1.01 \pm 0.03$\Rsun; which gives a corresponding surface
gravity of 
$\log g = 4.46 \pm 0.05$; in perfect agreement with the spectroscopic
value. 
We obtain an interval for the star's age between $0.12$ and $3.15$
Gyr.

%% Finally, using \citet{allen1973} to estimate the absolute V magnitude
%% and the color excess (taking the $E_{J-K}=0.0$ extinction into
%% account), we calculate the distance to the star to be 
%% $d=1\,310 \pm 100$ pc.
Finally, using simultaneously the seven BVr'i'JHKs broad-band magnitudes
reported in Table~\ref{startable} and following the method described
in \citet{gandolfi2008}, we derived an interstellar extinction
$A_{\mathrm V} = 0.20\pm0.10$~mag and a distance to the star
$d=1060\pm100$~pc. Consistent results were obtained using the absolute
magnitude \citep{allen1973} and the intrinsic near-infrared colours
\citep{gonzalezhernandez2009} tabulated for a G0 V type star.

\begin{table}
  \caption{Abundances of 10 elements in CoRoT-13 relative to the
    Sun. The element name, abundance relative to the Sun, and the
    number of spectral lines used are given.
  }
  \centering
  \begin{tabular}{*{2}{llr}}
    El.          & Abund.            & $N$ & El.          & Abund.         & $N$ \\       
    \hline
	{Li \sc i} & $ +1.45          $ &  1 & {Ti \sc  i} & $-0.06       $ &  2 \\
	{O  \sc i} & $ +0.02          $ &  2 & {Ti \sc ii} & $+0.05\pm0.04$ &  2 \\
	{Na \sc i} & $ -0.07          $ &  2 & {Cr \sc  i} & $-0.03\pm0.16$ &  4 \\
	{Mg \sc i} & $ -0.07          $ &  1 & {Fe \sc  i} & $+0.03\pm0.04$ &103 \\
	{Si \sc i} & $ +0.04  \pm 0.04$ & 13 & {Fe \sc ii} & $+0.04\pm0.05$ & 11 \\
	{Ca \sc i} & $ +0.03  \pm 0.05$ &  6 & {Ni \sc  i} & $-0.04\pm0.05$ & 25 \\
%%     {Li \sc   i} & $  1.70        $ &  1  & {Ti \sc   i} & $-0.01 \pm0.11$ &   4 \\
%%     {C  \sc   i} & $ -0.17        $ &  2  & {Ti \sc  ii} & $-0.06 \pm0.08$ &   3 \\
%%     {Na \sc   i} & $ -0.12 \pm0.07$ &  3  & {Cr \sc   i} & $-0.04 \pm0.11$ &   5 \\
%%     {Mg \sc   i} & $ -0.06        $ &  1  & {Fe \sc   i} & $-0.07 \pm0.07$ & 123 \\
%%     {Si \sc   i} & $ -0.06 \pm0.07$ &  8  & {Fe \sc  ii} & $-0.03 \pm0.08$ &  14 \\
%%     {Ca \sc   i} & $ -0.08 \pm0.07$ & 10  & {Ni \sc   i} & $-0.12 \pm0.07$ &  26 \\
%%     {Sc \sc  ii} & $ -0.12 \pm0.09$ &  3  & {Y  \sc  ii} & $-0.14 \pm0.12$ &   3 \\
  \end{tabular}
  \label{table:abund}      
\end{table}

%% \begin{figure}[t]
%%   \begin{center}
%%     \includegraphics[%
%%       angle=90,%
%%       width=0.9\linewidth,%
%%       height=0.5\textheight,%
%%       keepaspectratio]{HDHARPSCoRoT13b_vwaview}
%%   \end{center}
%%   \caption{
%%     Detail of the HARPS spectrum of CoRoT-13.
%%   }
%%   \label{fig:spec}
%% \end{figure}

%% \begin{figure}[t]
%%   \begin{center}
%%     \includegraphics[%
%%       angle=90,%
%%       width=0.9\linewidth,%
%%       height=0.5\textheight,%
%%       keepaspectratio]{HDHARPSCoRoT13b_vwaview_lithium}
%%   \end{center}
%%   \caption{
%%     Detail of the HARPS spectrum of CoRoT-13 in the region where Li
%%     was measured. 
%%   }
%%   \label{fig:spec2}
%% \end{figure}

%% \begin{figure}[t]
%%   \begin{center}
%%     \includegraphics[%
%%       angle=90,%
%%       width=0.9\linewidth,%
%%       height=0.5\textheight,%
%%       keepaspectratio]{vwa_pattern-corot13}
%% %      keepaspectratio]{corot13-vwa_pattern}
%%   \end{center}
%%   \caption{
%%     Mean abundances for 12 elements in CoRoT-13 HARPS
%%     spectrum. White circles correspond to neutral lines and red boxes
%%     stand for singly ionized lines. The yellow area represents the
%%     mean metallicity within one sigma error bar. 
%%   }
%%   \label{fig:abund}
%% \end{figure}

%............................................................................
\subsection{Radial Velocity measurements}

Precise radial velocity measurements of CoRoT-13 were obtained with
the HARPS spectrograph between the nights of November 22nd, 2009 and
February 15th, 2010 (ESO program 184.C-0639). 
HARPS is a cross-dispersed echelle spectrograph fiber-fed from the
Cassegrain focus of the 3.6 m telescope at La Silla Observatory, Chile
\citep{mayor2003}. 
Fifteen spectra with a spectral resolution $R \approx 115\,000$ were
obtained using exposure times of 3600 s, and setting one of the two
available fibers on the sky in order to monitor the presence of
moonlight and to obtain an optimal sky background subtraction, which
is important for faint targets such as this. 
The signal-to-noise ratios per pixel at 5500 \AA\, of these
observations range from 7.1 to 11.7. 
Th-Ar calibrations were obtained at the beginning of each night, which
has been shown to be enough to obtain the required precision, due to
the high stability of the instrument.

The spectra were reduced and extracted using the HARPS pipeline, and
the radial velocity was measured on each extracted spectrum by means
of a weighted cross-correlation \citep[see][]{baranne1996} with a
numerical mask corresponding to a G2 star. 
The resulting cross-correlation functions (CCFs) were fit by Gaussians
to get the radial velocities. 
The measured values are listed in Table~\ref{table:RV} and shown in
Fig.~\ref{fig:rv_vs_phase}, together with the best fit orbital
solution (see below). 
During some of the observations, the star fiber was contaminated by
moonlight. 
In those cases, if the peak of the CCF produced by moonlight was
expected to be close to the measured speed of the target, a correction
was applied using the fiber which recorded the sky (see Bonomo et
al. 2010, in preparation). 
Those points are shown as white circles in Fig.~\ref{fig:rv_vs_phase}
and we added quadratically 30\ms\, to the uncertainty estimated from
the CCF, in order to account for possible systematic errors introduced
by the moonlight correction.

% Hans comment 28.5
%The orbital solution was found by $\chi^2$ minimization, with the
%period and the time of passage through the periastron fixed to the
%values provided by the \corot\, ephemeris. 
The orbital solution was found by $\chi^2$ minimization, with the
period and epoch of inferior conjunction (when radial velocity is zero
after removal of the systemic velocity) being fixed to the values
provided by the \corot\, ephemeris 
(which are calculated by fitting a linear regression to the
center position of the individual transits).
The eccentricity of the orbit was a free parameter at first, but since
the best fit solution was compatible with a circular orbit at the
two-$\sigma$ level (the three-$\sigma$ upper limit to the eccentricity
is 0.145), we decided to fix it to $e=0$ for the determination of the
rest of the parameters and their uncertainties. 
Figure~\ref{fig:rv_vs_phase} shows the RV measurements, phased to the
\corot\, period, together with the best fit circular model and the
residuals; the obtained parameters are listed in
Table~\ref{starplanet_param_table}.
The resulting value of $\chi^2$ is 8.2 for 13 degrees of freedom, and
the $rms$ of the residuals is $20.2\:\mathrm{m\,s^{-1}}$, which is
compatible with what should be expected based on the median of the RV
uncertainties, $21.2\:\mathrm{m\,s^{-1}}$. 
These facts suggest that the circular model -- with the obtained
parameters-- adequately describes the available data. 
                %Indeed, the value of the reduced $chi^2$ statistics
		%being below unity means that errors have been
		%overestimated. Without inflating the error of the
		%points corrected by the moon contamination, we obtain
		%similar results in the fit,  

With fixed ephemeris and eccentricity set to zero, the fitting problem 
is reduced to a linear least-square minimization with two free
parameters. 
The uncertainties reported in Table~\ref{starplanet_param_table} are
therefore estimated by means of the covariance matrix, which has a
covariance term of $6.48\;\mathrm{m^2\,s^{-2}}$. 
However, stellar activity and other long-term phenomena can produce
correlated noise in the observations and hence render the above
estimation of the uncertainties invalid. 
In order to explore this we used the Prayer Bead method 
\citep[see, for example,][]{desert2009,winn2009}, i.e. we performed a
cyclic permutation of the residuals of the best fit curve and fit the
model again. 
We repeated this for every possible shift and measured the standard
deviation of the obtained parameters. 
We also performed a similar analysis but randomly re-ordering the
residuals rather than shifting them. In this way, we constructed 10000
synthetic data sets that were used to fit our model again. 
In both cases, the obtained dispersion of the parameters were smaller
than the error bars reported in Table~\ref{starplanet_param_table}. 

The bisector analysis for these data is shown in
Fig~\ref{bisanalysis}, where the uncertainty in the bisector span
velocity has been set to twice that of the corresponding radial
velocity. 
The bisector span velocities do not show any clear dependence with
radial velocity values and the Pearson correlation coefficient between
these two magnitudes is around 0.15, which is a sign of lack of
correlation. 
This fact clearly indicates that the measured RV variations do not
originate from changes in the shape in the CCF as would be the case if
the system consisted of a background eclipsing binary whose light were
diluted by the \corot\, main target. 
Additionally, no significant changes in the measured RV are observed
when different stellar masks are used for the correlation, which
further excludes the background eclipsing binary scenario.
%which also favors the planetary scenario over the background eclipsing binary one.  
 
\begin{table}
\begin{center}
\caption{Radial velocities measurements and bisector span velocities measured with HARPS.}
\begin{tabular}{lllll}
\hline\hline
BJD&RV&$\sigma_\mathrm{RV}$&Bis&Moon Correction\tablefootmark{a}\\
-2455000&[km/s]&[km/s]&[km/s]&[km/s]\\
\hline
158.7187&22.293&0.029&0.027&\\
160.725&22.598&0.026&0.038&\\
161.7209&22.522&0.053&0.144&0.101\\
163.8279&22.379&0.044&-0.019&0.055\\
166.7245&22.311&0.037&-0.091&0.219\\
168.7634&22.641&0.042&-0.012&0.503\\
219.7315&22.296&0.02&0.036&\\
220.7327&22.464&0.022&-0.014&\\
225.6511&22.606&0.021&-0.01&\\
226.7338&22.45&0.021&0.059&\\
228.6545&22.452&0.021&-0&\\
237.5841&22.624&0.02&0.006&\\
238.5617&22.478&0.018&-0.014&\\
239.7142&22.304&0.019&0.024&\\
243.6139&22.311&0.016&-0.003&\\
\hline
\end{tabular}
\label{table:RV}
\end{center}
\tablefoottext{a}{Difference between moonlight-corrected and
  uncorrected radial velocities.}
\end{table}

%\begin{figure}[t]
%  \begin{center}
%    \includegraphics[%
%      width=0.9\linewidth,%
%      height=0.5\textheight,%
%      keepaspectratio]{rv_vs_time}
%  \end{center}
%  \caption{
%  \label{fig:rv_vs_t}
%   Radial Velocity measurements taken with HARPS on CoRoT-13 and
%    its orbital solution.}
%   \end{figure}

\begin{figure}[t]
  \begin{center}
    \includegraphics[%
      width=0.9\linewidth,%
      height=0.5\textheight,%
%      bb=11 173 600 618,clip,%
      viewport=11 173 600 618,%
      keepaspectratio]{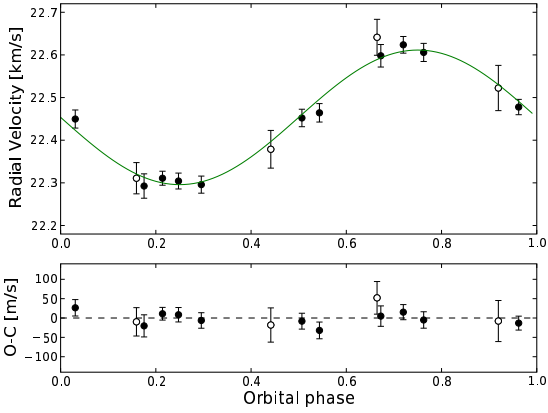}
  \end{center}
  \caption{
  \label{fig:rv_vs_phase}
  \emph{Upper panel}: Radial Velocity measurements phased to the
  orbital period measured by \corot. The solid curve represents the
  best fit solution. \emph{Lower panel}: Residuals to the fit. In both
  panels, the white symbols indicate the measurements that have been
  corrected for moonlight contamination. 
  }
  \end{figure}

\begin{figure}[t]
  \begin{center}
    \includegraphics[%
      width=0.9\linewidth,%
      height=0.5\textheight,%
%      bb=11 173 600 618,clip,%
      viewport=11 173 600 618,%
      keepaspectratio]{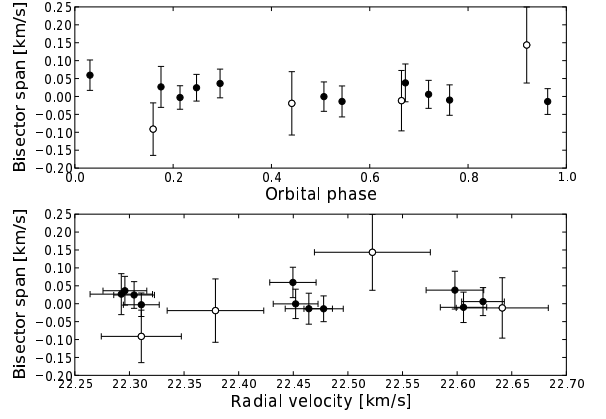}
  \end{center}
  \caption{
    Bisector analysis of the HARPS data. The white symbols represent
    measurements that have been corrected for moonlight contamination. 
  }
  \label{bisanalysis}
\end{figure}

%\begin{figure}[t]
%  \begin{center}
%    \includegraphics[%
%      width=0.9\linewidth,%
%      height=0.5\textheight,%
%      keepaspectratio]{orbit_oc_allorders_rescerr}
%  \end{center}
%  \caption{
%    Radial Velocity measurements taken with HARPS on CoRoT-13 and
%    its orbital solution.
%  }
%  \label{fig:harps}
%\end{figure}

%\begin{figure}[t]
%  \begin{center}
%    \includegraphics[%
%      width=0.9\linewidth,%
%      height=0.5\textheight,%
%      keepaspectratio]{discarded_planets}
%  \end{center}
%  \caption{
%    Parameter space of the discarded additional planets in the system
%    with current HARPS measurements.
%  }
%  \label{fig:rvdrift}
%\end{figure}

%
%____________________________________________________________________________
\section{Planetary parameters}
\label{sec:planetary_parameters}

The transit parameters were determined in the same manner as described
in the case of CoRoT-6b \citep{fridlund2010}. 
The pre-processed light curve was divided by its median, and a new
light curve was constructed by convolving it with a fourth order
Savitzky-Golay filter \citep{press2002}. 
The standard deviation of the differences between the measured and the
convolved light curves is calculated and a 5$\sigma$ clip is applied
to remove spurious outliers. 
This clipping is iterated until no more outliers are found. 
After these steps, the light curve is folded and a bin average is
applied forming 5000 bins.

We used the \citet{mandel2002} model to perform the fit to the final,
phase-folded light curve. 
We assumed a circular orbit and an absolutely dark planet at the CoRoT
wavelength. 
% which does not contribute to the observed light. 
%% Our six adjusted parameters were: the $a/R_{\rm s}$, 
%% $k = R_{\rm pl}/R_{\rm s}$ ratios, the impact parameter 
%% $b=a*\cos i/R_{\rm s}$ ($i$ is the inclination of the orbit) and the
%% two limb darkening coefficients $u_{+}$, $u_{-}$ (a quadratic limb
%% darkening law was used). 
%% The sixth parameter was the contaminant factor which was varied
%% between 10-12\% (corresponding to the measured value of $11\pm1$\%) to
%% allow a better estimation of errors.

The six adjusted parameters were: the $a/R_{\rm s}$, 
$k = R_{\rm pl}/R_{\rm s}$ ratios, the impact parameter 
$b=a*\cos i/R_{\rm s}$ ($i$ is the inclination of the orbit), the
two limb darkening coefficients $u_{+}$, $u_{-}$ and the contaminant
factor, which was varied between 10-12\% (corresponding to the
measured value of $11\pm1$\%) to allow a better estimation of errors.
The quadratic limb darkening law is 
$I(\mu)/I(1)= 1-\mu+u_{a}\mu + u_{b} (1-\mu)^2;$
and we fit the coefficients 
$u_{+}=u_{a}+u_{b}$ and
$u_{-}=u_{a}-u_{b}$. 
The best fit was found by the Harmony Search algorithm
\citep{geem2001}, a genetic-type algorithm. 
The $1\sigma$ errors were obtained from the width of the parameter
distribution to be between $\chi^2_{\rm min}$ and 
$\chi^2_{\rm min} + 1$.  

We calculated different solutions for the light curve of CoRoT-13b
using different approaches in the treatment of the limb darkening
parameters in order to understand the constraints that this effect
puts in the determination of the final values. 
The results are presented in Table~\ref{table:szilardsolutions}. 
In model A we left both $u_{+}$ and $u_{-}$ to be free parameters. 
In models B, C and D we fixed $u_{+}$ at 0.81 (the value what we found
in model A), at 0.88 (i.e. $0.81+1\sigma$) and 0.74 ($0.81-1\sigma$;
where $\sigma$ means the error bar of the coefficient found in model
A) respectively. 
In model E we fixed the limb darkening to the predicted values for the
star \citep{sing2010}. 
The corresponding $\chi^2$-values of these solutions were not too
different from each other, showing that limb darkening is a second
order effect in the transit shape. 
%One can see that 
The different assumptions for the limb-darkening yielded quite
consistent light curve solutions: the $a/R_{\rm s}$, $k$ ratios are in 
good agreement well within their error bars in these five models. 
When we fixed $u_{+}$, the uncertainties of the impact parameter
increased by a factor of two, however, the error bars of  impact
parameters found in the B, C, D and E solutions are overlapping with
the one found in  model A. 
The precise determination of the impact parameter is more complicated
when the transit is nearly central. 
We learn from this experiment that fitting both limb darkening
coefficients (or, what we actually did, fitting both of their
combinations) yielded a bit more precise values.

\begin{figure}[t]
  \begin{center}
    \includegraphics[%
      width=0.9\linewidth,%
      height=0.5\textheight,%
%      bb=54 360 558 720, clip,%
      viewport=54 360 558 720,%
      keepaspectratio]{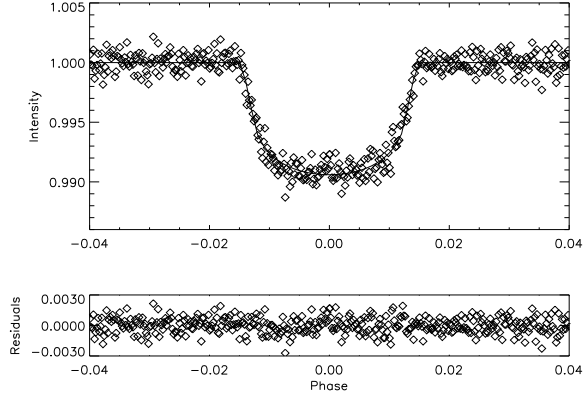}
  \end{center}
  \caption{
    Folded transit of CoRoT-13b, best fit and its residuals using
    the values of solution A in Table~\ref{table:szilardsolutions}.
  }
  \label{fig:c13b_fit}
\end{figure}

Since $u_{+} = 0.81 \pm 0.07$ (\citealt{sing2010} gives 
$0.662 \pm 0.022$ for this temperature, log g and metallicity) and 
$u_{-} = -0.09 \pm 0.09$ (\citealt{sing2010} gives $0.156 \pm 0.022$)
the agreement between theoretical predictions and measurements is
withing 2 sigma error bars, which is satisfactory. 
%% This seems to be quite satisfactory at this stage of the
%% theory. 
%% Interestingly, solution D is closer to the theoretical predictions,  
% Hans comment 28.05.2010
%but nothing else makes think that this better
%agreement would say that it is the better solution.
%% but there are no further indications that this better agreement
%% constitutes also a more correct description of the system.
%All solutions produces good agreement (within 2-sigma error bars) with
%the theoretical predictions. 
We accept solution A as the definitive one (the fit is shown in
Fig.~\ref{fig:c13b_fit}). 

\begin{table*}
\caption{Values of the adjusted parameters in the modeling of the
  transit of CoRoT-13b in the different approximations described in
  the text.}
\centering
\begin{tabular}{*{6}{c}}
\hline\hline
Parameter                         & A                         & B                         & C                         & D                         & E                         \\
\hline
$a/R_{\mathrm{*}}$                & $   10.81 \pm 0.32      $ & $   11.22 \pm 0.42      $ & $   11.15 \pm 0.40      $ & $   10.98 \pm 0.48      $ & $  10.89 \pm 0.40       $ \\
$k=R_{\mathrm{p}}/R_{\mathrm{*}}$ & $  0.0909 \pm 0.0014    $ & $  0.0900 \pm 0.0016    $ & $  0.0896 \pm 0.0014    $ & $  0.0912 \pm 0.0012    $ & $ 0.0914 \pm 0.0011     $ \\
$b=a\,\cos\,i/R_{\mathrm{*}}$     & $   0.374 \pm 0.054     $ & $   0.264 \pm 0.097     $ & $   0.271 \pm 0.092     $ & $   0.349 \pm 0.091     $ & $  0.385 \pm 0.070      $ \\
inclination (deg)                 & $ 88.01^{+0.35}_{-0.33} $ & $ 88.65^{+0.57}_{-0.53} $ & $ 88.61^{+0.54}_{-0.50} $ & $ 88.01^{+0.58}_{-0.53} $ & $ 87.97^{+0.43}_{-0.46} $ \\
$u_{+}$                           & $    0.81 \pm 0.07      $ & $  0.81$ (fixed)          & $ 0.88$ (fixed)           & $ 0.74$ (fixed)           & $ 0.662$ (fixed)          \\ 
$u_{-}$                           & $   -0.09 \pm 0.09      $ & $ -0.09$ (fixed)          & $   -0.24 \pm 0.32      $ & $   -0.01 \pm 0.25      $ & $ 0.156$ (fixed)          \\
third light                       & $    0.11 \pm 0.01      $ & $    0.11 \pm 0.01      $ & $    0.11 \pm 0.01      $ & $    0.11 \pm 0.01      $ & $    0.11 \pm 0.01      $ \\
$\chi^2$\tablefootmark{(a)}       & $ 1.00000               $ & $ 1.00039               $ & $ 0.99945               $ & $ 1.00077               $ & $ 1.00130               $ \\
\end{tabular}
\tablefoot{
  \tablefoottext{a}{Normalized to the value of solution A}
}
\label{table:szilardsolutions}      
\end{table*}

% table IDs, coordinates and magnitudes (first version)
% \begin{table}[h]
% \caption{ IDs, coordinates and magnitudes.}            
% \centering        
% \begin{minipage}[!]{7.0cm}  
% \renewcommand{\footnoterule}{}     
% \begin{tabular}{lcc}       
% \hline\hline                 
% CoRoT window ID & \\
% CoRoT ID & \\
% USNO-A2 ID  & \\
% 2MASS ID   &  \\
% GSC2.3 ID &  \\
% \\
% \multicolumn{2}{l}{Coordinates} \\
% \hline            
% RA (J2000)  &  \\
% Dec (J2000) &  \\
% \\
% \multicolumn{3}{l}{Magnitudes} \\
% \hline
% \centering
% Filter & Mag & Error \\
% \hline
% B$^a$  & & \\
% V$^a$  & & \\
% r'$^a$ & & \\
% i'$^a$ & & \\
% J$^b$  & & \\
% H$^b$  & & \\
% K$^b$  & & \\
% %\\                                    
% %\multicolumn{3}{l}{Proper motion} \\
% %\hline
% %$\mu_{\alpha}$ & 8.0   & ''/yr
% %$\mu_{\alpha}$ & -11.8 & ''/yr
% \hline\hline
% \vspace{-0.5cm}
% \footnotetext[1]{Provided by Exo-Dat (Deleuil et al, 2008);}
% \footnotetext[2]{from 2MASS catalog.}
% \end{tabular}
% \end{minipage}
% \label{startable}      
% \end{table}

% table Planet and star parameters
\begin{table}
%\centering
\caption{Planet and star parameters.}            
\begin{tabular}{l l}
\hline\hline                 
\multicolumn{2}{l}{\emph{Ephemeris}} \\
\hline
Planet orbital period $P$ [days]                         & $    4.035\,190 \pm 0.000\,030$ \\ % $4.035\,189\,9 \pm 0.000\,030\,26$
Primary transit epoch $T_\mathrm{tr}$ [HJD-2\,450\,000]  & $ 4\,790.809\,1 \pm 0.000\,6$   \\
Primary transit duration $d_\mathrm{tr}$ [h]             & $          3.14 \pm 0.01 $      \\
%% add the following if the secondary eclipse is detected or if the orbit is eccentric and eclipse timing can be predicted
%% Secondary transit epoch $T_\mathrm{*}$ [HJD-2\,400\,000] &  $\pm$  \\
%% Secondary transit duration $d_\mathrm{*}$ [h]            &  $\pm$  \\
& \\
\multicolumn{2}{l}{\emph{Results from radial velocity observations}} \\
\hline    
%Epoch of periastron $T_{0}$ [HJD-2\,400\,000] &  $        \pm       $  \\
Orbital eccentricity $e$                      &  0 (fixed)             \\
%Argument of periastron $\omega$ [deg]         &  $        \pm       $  \\ 
Radial velocity semi-amplitude $K$ [\ms]      &  $  157.8 \pm 7.7   $  \\
Systemic velocity $V_{r}$ [\kms]              &  $22.4536 \pm 0.0060$  \\
O-C residuals [\ms]                           &  $ 20.2             $  \\
& \\
\multicolumn{2}{l}{\emph{Fitted transit parameters}} \\
\hline
%Radius ratio $k=R_\mathrm{p}/R_{*}$                            & $0.0909 \pm 0.0014$ \\
%Linear limb darkening coefficient\tablefootmark{a} $u$         & $0.36   \pm 0.08$   \\
%Impact parameter\tablefootmark{b} $b$                          &  $0.374 \pm 0.054$  \\
Scaled semi-major axis $a/R_{*}$                               & $   10.81 \pm 0.32    $ \\
Radius ratio $k=R_\mathrm{p}/R_{*}$                            & $0.090\,9 \pm 0.001\,4$ \\
Quadratic limb darkening coefficients\tablefootmark{a} $u_{+}$ & $    0.81 \pm 0.07    $ \\
\phantom{Quadratic limb darkening coefficients\,}      $u_{-}$ & $   -0.09 \pm 0.09    $ \\
Impact parameter\tablefootmark{b} $b$                          & $   0.374 \pm 0.054   $ \\
%Depth of secondary eclipse                                     &  $\pm$  \\
& \\
\multicolumn{2}{l}{\emph{Deduced transit parameters}} \\
\hline
% footnote IF orbit is eccentric
%Scaled semi-major axis$\tablefootmark{c}           &  $\pm$  \\
$M^{1/3}_{*}/R_{*}$ [solar units]                  & $1.014 \pm 0.030$  \\
Stellar density $\rho_{*}$ [\gcm3]                 & $1.468 \pm 0.131$  \\
%Inclination $i$ [deg]                              & $88.01^{+0.35}_{-0.33}$  \\
Inclination $i$ [deg]                              & $88.02^{+0.34}_{-0.36}$  \\
& \\
\multicolumn{2}{l}{\emph{Spectroscopic parameters }} \\
\hline
Effective temperature $T_\mathrm{eff}$ [K]  & $5\,945 \pm 90  $ \\
Surface gravity log\,$g$ [dex]              & $  4.30 \pm 0.10$ \\
Metallicity $[\rm{Fe/H}]$ [dex]             & $  0.01 \pm 0.07$ \\
Stellar rotational velocity {\vsini} [\kms] & $     4 \pm 1   $ \\
Spectral type                               & G0V               \\
& \\
\multicolumn{2}{l}{\emph{Stellar and planetary physical parameters from combined analysis}} \\
\hline
Star mass [\Msun]                                                & $    1.09 \pm 0.02   $ \\
Star radius [\Rsun]                                              & $    1.01 \pm 0.03   $ \\
Surface gravity log\,$g$ [dex]                                   & $    4.46 \pm 0.05   $ \\
Age of the star $t$ [Gyr]                                        & $     0.12 - 3.15    $ \\
Distance of the system [pc]                                      & $  1\,060 \pm 100    $ \\
interstellar extinction $A_{\mathrm V}$ [mag]                    & $     0.20\pm 0.10   $ \\
Stellar rotation period $P_\mathrm{rot}$ [days]                  & $ 13^{+5}_{-3}       $ \\
Orbital semi-major axis $a$ [AU]                                 & $  0.0510 \pm 0.0031 $ \\
%% add following if the planetary orbit is excentric
% Orbital distance at periastron $a_\mathrm{per}$ [AU] &  $\pm$  \\
% Orbital distance at apastron $a_\mathrm{apo}$ [AU]   &  $\pm$  \\
Planet mass $M_\mathrm{p}$ [\Mjup]\tablefootmark{c}              & $   1.308 \pm 0.066  $ \\
Planet radius $R_\mathrm{p}$ [\Rjup]\tablefootmark{c}            & $   0.885 \pm 0.014  $ \\
Planet density $\rho_\mathrm{p}$ [\gcm3]                         & $    2.34 \pm 0.23   $ \\
Planet surface gravity log\,$g$ [dex]                            & $    3.62 \pm 0.03   $ \\
Average surface temperature\tablefootmark{d} $T_{\mathrm p}$ [K] & $    \sim 1\,700     $ \\
%Equilibrium temperature\tablefootmark{e} $T^\mathrm{per}_\mathrm{eq}$ [K] &  $\sim 1\,700$  \\
% Equilibrium temperature at periastron\tablefootmark{e} $T^\mathrm{per}_\mathrm{eq}$ [K] & $\pm$ \\
% Equilibrium temperature at apastron\tablefootmark{e} $T^\mathrm{apo}_\mathrm{eq}$ [K] & $\pm$ \\
\hline       
\end{tabular}
\tablefoot{
  \tablefoottext{a}{$I(\mu)/I(1)= 1-\mu+u_{a}\mu + u_{b} (1-\mu)^2$,
  where $I(1)$ is the specific intensity at the center of the disk and
  $\mu=\cos{\gamma}$, $\gamma$ being the angle between the surface
  normal and the line of sight; $u_{+}=u_{a}+u_{b}$ and
  $u_{-}=u_{a}-u_{b}$.}
  \tablefoottext{b}{$b=\frac{a \cdot \cos{i}}{R_{*}}$}
  % \tablefoottext{b}{$b=\frac{a \cdot \cos{i}}{R_{*}} \cdot \frac{1-e^{2}}{1+e \cdot \sin{\omega}}$}
  % next footnote IF orbit is eccentric
%  \tablefoottext{c}{$a/R_{*}=\frac{1+e \cdot \cos{\nu_{1}}}{1-e^{2}} \cdot \frac{1+k}{\sqrt{1-\cos^{2}({\nu_{1}+\omega-\frac{\pi}{2}}) \cdot \sin^{2}{i}}}$, where $\nu_{1}$ is the true anomaly measured from the periastron passage at the transit egress (see Gim09) .}
  \tablefoottext{c}{Radius and mass of Jupiter taken as $71\,492$ km
  and $1.8992 \times 10^{30}$ g, respectively \citep{lang1999}.}
  \tablefoottext{d}{Zero albedo equilibrium temperature for an
  isotropic planetary emission.}
}
\label{starplanet_param_table}  
\end{table}

\section{Discussion}
\label{sec:discussion}

%............................................................................
\subsection{Stellar properties}
\label{subsec:stellar_properties}

%% \begin{figure}[t]
%%   \begin{center}
%%     \includegraphics[%
%%       width=0.9\linewidth,%
%%       height=0.5\textheight,%
%%       keepaspectratio]{CoRoT_13b_lomb01}
%%   \end{center}
%%   \caption{
%%     Lomb-Scargle periodogram of the raw LC corrected only from the
%%     jumps produced by hot-pixels.  
%%   }
%%   \label{fig:CoRoT_13b_lomb01}
%% \end{figure}

%% \begin{figure}[t]
%%   \begin{center}
%%     \includegraphics[%
%%       width=0.9\linewidth,%
%%       height=0.5\textheight,%
%%       keepaspectratio]{CoRoT_13b_lomb04-lite}
%%   \end{center}
%%   \caption{
%%     LC of CoRoT-13b with the best harmonic fit and the LC of its
%%     closest contaminant.
%%   }
%%   \label{fig:CoRoT_13b_lomb04}
%% \end{figure}

%\begin{figure}[t]
%  \begin{center}
%    \includegraphics[%
%      width=0.9\linewidth,%
%      height=0.5\textheight,%
%      keepaspectratio]{CoRoT_13b_lomb05}
%  \end{center}
%  \caption{
%    Lomb-Scargle periodogram of the raw LC of CoRoT-13b and its
%    closest contaminant.
%  }
%  \label{fig:CoRoT_13b_lomb05}
%\end{figure}

The spectroscopic analysis of CoRoT-13 reveals a G0V star with an age
between $0.12$ and $3.15$ Gyr, solar metallicity 
(${\rm [M/H]} = +0.01 \pm 0.07$), a high relative abundance of lithium 
($+1.45$ dex), and a low activity level according to the analysis of
activity indicators such as the H-K {Ca \sc ii} lines 
(where no emission is detected).

The rate of lithium depletion of solar like stars is related to the
age of the star and to the depth of the convective zone, as it is
destroyed at a temperature of approximately $\sim 2.5 \cdot 10^6$\,K  
in the radiative region of a star \citep{chaboyer1998}. 
Given the spectral type of CoRoT-13, we expect a lower lithium
depletion rate than in solar analogs \citep{castro2009}. 
Using the value of the {Li \sc i} abundance, we compute a 
$\log n(Li) = 2.55$. 
From Fig.~7 of \citealt{sestito2005} and with the value of the
effective temperature  ($T_{\rm eff} = 5945 \pm 90$\,K) we estimate
the age of the star in the range $300$ Myr to $1$ Gyr, consistent with
the range from the evolutionary models. 
In a recent paper, \citet{israelian2009} claim a lithium depletion in
solar like stars with orbiting planets, although it is not clear that
previous observations support this conclusion \citep{melendez2009b}. 
CoRoT-13 is not depleted in lithium, albeit we call the attention to
the fact that the effective temperature of this star is slightly
higher than the upper limit for depletion given in
\citet{israelian2009}.

The \vsini\, value indicates a rotational period of the star of around
13 days\footnote{a lower limit, as the value of $\sin\,i$ for the spin
  axis of the star is unknown.}. 
Gyrochronology \citep{barnes2007} could be used as an age
estimator. Using the improved gyrochronology relations from
\citet{mamajek2008} we derived a gyrochronologic age of 1.66 Gyr well
within the range of age given by evolutionary models. No emission
feature is seen  at the bottom of the \ion{Ca}{ii} H and K lines nor
in the H$_\alpha$ line showing that the star belongs to the inactive 
population with $\log~R^\prime_HK < -5.0$. We thus didn't derive any 
chromospheric age for the star.
%% According to this value, a $(B-V)_{0}=0.60$; gyro-chronologically
%% predicts an age of $1.4$ Gyr \citep{barnes2007} which is in agreement
%% within the  
%% %very large 
%% range of age given by evolutionary models. 
%% The latter paper recalls an equation that links the age and the
%% activity level and the expected value of 
%% $\log R'_{HK}$ is $\sim -6.0$; which is very low and consistent with
%% our observations. 
Other G0V stars are found with similar rotation rates and low activity
levels \citep{noyes1984}.

We have looked for signs of stellar rotation in the light
curve (LC) observed by \corot\, to make a comparison with the clear
signs of spot modulation found in the cases of CoRoT-2b
\citep{lanza2009a}, CoRoT-4b \citep{lanza2009b}, CoRoT-6b
\citep{fridlund2010} or CoRoT-7b \citep{lanza2010b}. 
The Lomb-Scargle periodogram of the LC, once the planetary transits
have been removed and the hot-pixel events have been treated, shows
indeed a significant broad peak around 77 days; but not any
significant peak at the expected rotational frequencies.
The 77 days period is comparable with the length of the run (115
days), so it might be that we are observing an irregular pattern that,
in the observing window of 115 days, has a typical timescale of
variation of 77 days that may mimic an harmonic oscillation. 
The only reliable information on the nature of this particular signal
is its characteristic amplitude of 0.5\% and its characteristic
timescale of 77 days; one has to be extremely cautious when
interpreting its nature. 
Two immediate possibilities are stellar activity and an instrumental
residual signal. 
In the latter case, stray light or other instrumental effects such as
temperature fluctuations should affect a region of the CCD (if not
all), instead of a single target. 
None of the targets in the neighborhood of CoRoT-13 show a similar
pattern. 
%(see Fig.~\ref{fig:CoRoT_13b_lomb05}).
Other environmental features (such as hot pixels) although present,
have a completely different behavior both in amplitude and in
timescale and are not likely to be responsible for the signal. 
% Hans comment 28.5.2010
The spectroscopic analysis shows that CoRoT-13 is a quiet dwarf star. 
A G0V star might show spot modulation with a characteristic timescale
similar to the rotational period of the star as well as long term
variations in timescales of several years \citep{baliunas1995}, but
not in the timescales considered here.
On the other hand, slow rotating giant stars might show a pattern of
variability with similar amplitudes and characteristic timescales as 
the ones revealed by the periodogram.
We therefore conclude that the modulation measured in the LC of
\corot-13 is due to a background contaminant and not to the main
target.

In a recent paper, \citet{lanza2010a} links the presence of giant 
planets and the angular momentum loss of their host stars, detecting
that giant planet hosting stars with $T_{\rm eff} \gtrsim 6000 K$ tend
to be in a $n/\Omega \approx 1,2$ orbital period/rotational period
synchronization with their respective planets. The effective
temperature of CoRoT-13 lays right below this lower limit and indeed
the rotational period obtained from the spectroscopic analysis (around
13 days) is longer than the expected 2:1 resonance or the
synchronization; although the rotational period is not sufficiently
well constrained (the relative uncertainty is around 35\%) and the
inclination of the spin-axis of the star is not known. A study of the
Rossiter-McLaughlin effect of this planet will provide the relative
angle of inclination of the planet and the star, which is an important
parameter in the model proposed by \citet{lanza2010a}.

%............................................................................
\subsection{Planet interior}
\label{subsec:planet_interior}

\begin{figure}[t]
  \begin{center}
    \includegraphics[%
      width=0.9\linewidth,%
      height=0.5\textheight,%
      viewport=50 50 410 302,%
      keepaspectratio]{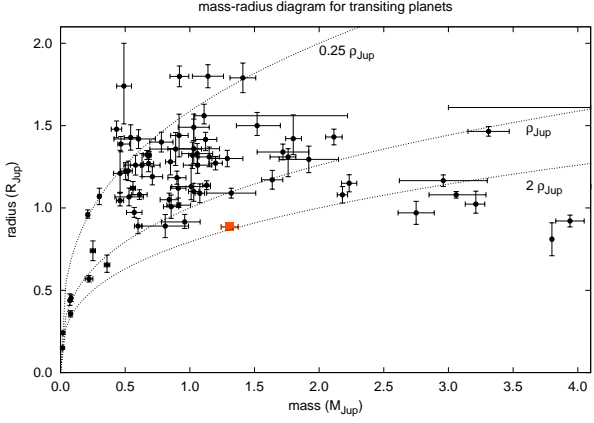}
  \end{center}
  \caption{
    Position of CoRoT-13b (square) among the other transiting
    planets in a mass-radius diagram. 
  }
  \label{fig:mass_radius}
\end{figure}

\begin{figure}[t]
  \begin{center}
    \includegraphics[%
      width=0.9\linewidth,%
      height=0.5\textheight,%
      keepaspectratio]{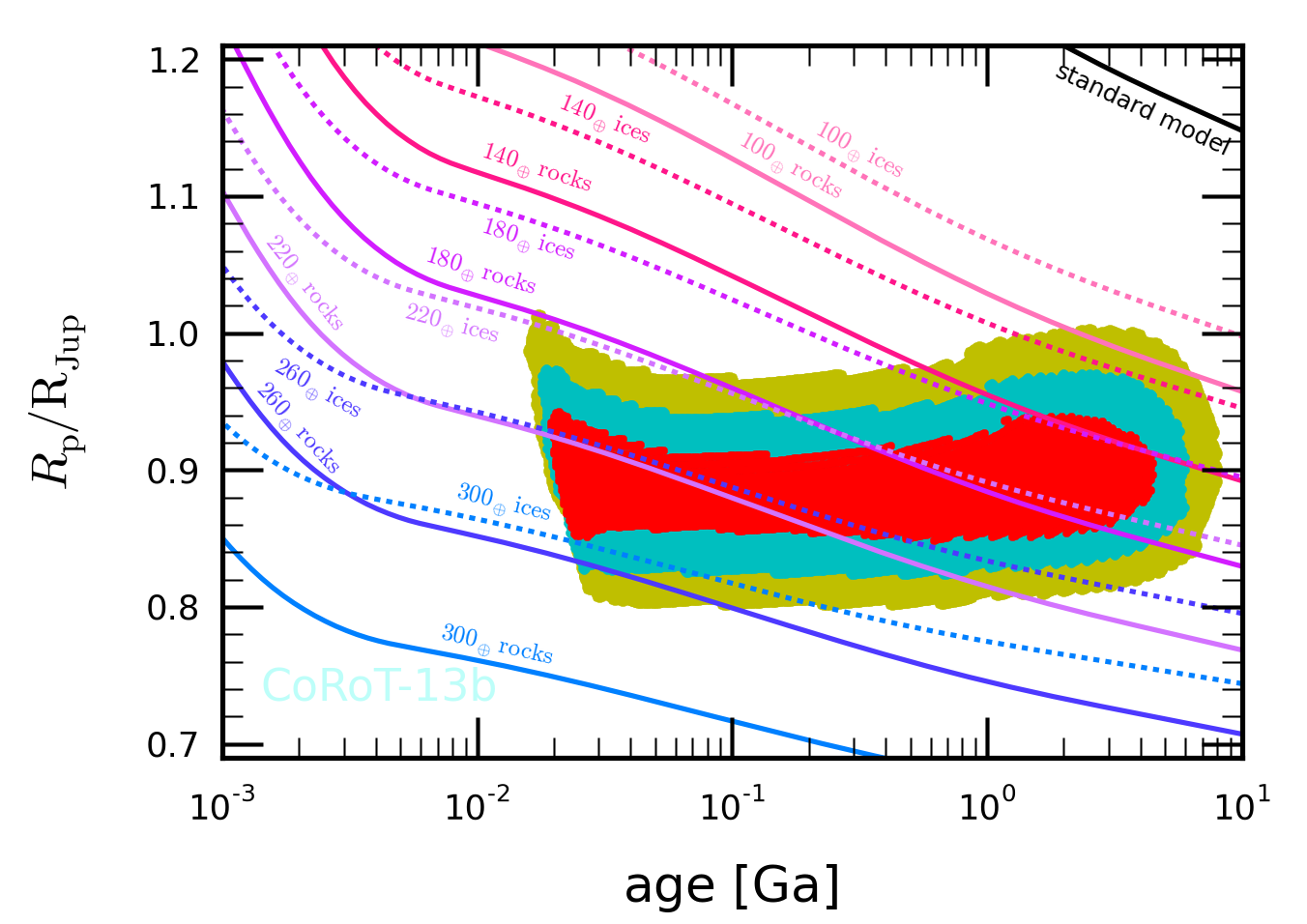}
  \end{center}
  \caption{Age (in Ga=$10^9\,$years) versus transit radius of
    CoRoT-13b (in Jupiter units, $1$\Rjup=$71\,492$\,km). The
    colored area correspond to constraints derived from stellar
    evolution models matching the stellar density and effective
    temperature within a certain number of standard deviations: less
    than $1\sigma$ (red), $2\sigma$ (blue) or $3\sigma$ (yellow). The
    curves are evolution tracks for CoRoT-13b (assuming
    $M=1.308$\Mjup, $T_{\rm eq}=1700\,$K), with various models
    as labelled 
  }
  \label{fig:internal_structure}
\end{figure}

%% Figure~\ref{fig:mass_radius} shows the position of CoRoT-13b in a 
%% mass-radius diagram among other transiting planets. Its density is
%% relatively higher than other hot jupiters in the same region of the
%% mass/radius diagram, although it is not exceptionally dense. 
Figure~\ref{fig:mass_radius} compares CoRoT-13b to other transiting
exoplanets in a mass-radius diagram. 
Although it is not the densest object known so far (both super-Earths
and brown dwarfs may be denser), it is clearly extremely dense for its
mass. 
This is confirmed by a combined modeling of the star and planet
evolution \citep[see][]{borde2010,morel2008,guillot1995} shown in
Fig.~\ref{fig:internal_structure}. 
The small planetary radius derived from the transit photometry and
spectroscopy can be explained only by advocating the presence of a
truly considerable amount of heavy elements in the planet. 
When fitting the stellar effective temperature and density within
their 1-sigma error bars (see Table~\ref{starplanet_param_table} and
red area in Fig.~\ref{fig:internal_structure}), between about $140$
and $300$\Mearth\, of heavy elements are required to reproduce the
measured planetary size. 
When fitting the stellar parameters only within $3\sigma$ (yellow
area), still at least $100$\Mearth\, of heavy elements are needed. 
In our planetary evolutions calculations, we assumed all heavy
elements to be grouped into a well-defined central core, surrounded by
a solar-composition envelope, with no added sources of heat. 
The possibility that these heavy elements may be at least partly mixed
in the envelope is not expected to change these numbers significantly
\citep{guillot2005,ikoma2006,baraffe2008}.

This extremely high amount of heavy elements is surprising for several
reasons: first, this is probably a record - HD149026b, a Saturn-mass
planet, has been known to possess about $60-70$\Mearth\, of heavy
elements \citep{sato2005,ikoma2006}. 
Other objects, such as OGLE-TR-56b and OGLE-TR-132b were also later
shown to have close to $100$\Mearth\, in heavy elements
\citep{guillot2006}, but no planetary-mass object was yet shown to
possess more heavy elements. 
An exception may be HAT-P-2b, initially thought to have $M_{Z}=200$ to
$600$\Mearth\, \citep{baraffe2008,leconte2009}, but a revision of the 
stellar parameters \citep{torres2008} yields much smaller $M_Z$ values
for this object.  

Second, planet formation models do not predict the existence of such
dense objects \citep{mordasini2009}. 
This is because a protoplanetary core growing beyond a few tens of
Earth masses rapidly captures any surrounding hydrogen and helium, and
that the accretion of planetesimals is suppressed by the growth of the
protoplanet beyond Saturn's mass. 
Giant impacts need to be invoked for a further capture of a
significant mass of heavy elements \citep{ikoma2006}.

Third, CoRoT-13b is a counterexample outside the correlation between
stellar-metallicity and planetary mass in heavy elements
\citep{guillot2006,burrows2007,guillot2008}. 
One interesting possibility however is related to the high lithium
abundance of the star. 
If the metallicity of stars with planets in fact tell us about the
late accretion of circumstellar gas filtered of their heavy elements
by planet formation \citep{melendez2009b,ramirez2009,nordlund2009}, it
raises the possibility that the CoRoT-13 system was in fact
metal-rich, but that a heavy-elements poor, lithium-rich rare last
burst of accretion modified the chemical properties of the star's thin
outer convective zone.

%............................................................................
\subsection{Thermal losses}
\label{subsec:thermal_losses}

We estimate the thermal mass loss of CoRoT-13b by using the method and
formulae described in \citet{lammer2009}, and we find a negligible 
escape rate which did not influence the planets mass over its 
history. The reason of the negligible thermal mass loss of CoRoT-13b
is the planets compactness and high density of about
$2.34$\gcm3.

%............................................................................
\subsection{Occultation of the planet by the star}
\label{subsec:secondary_transit}

\begin{figure}[t]
  \begin{center}
    \includegraphics[%
      width=0.9\linewidth,%
      height=0.5\textheight,%
%      bb=54 360 558 720,clip,%
      viewport=54 360 558 720,%
      keepaspectratio]{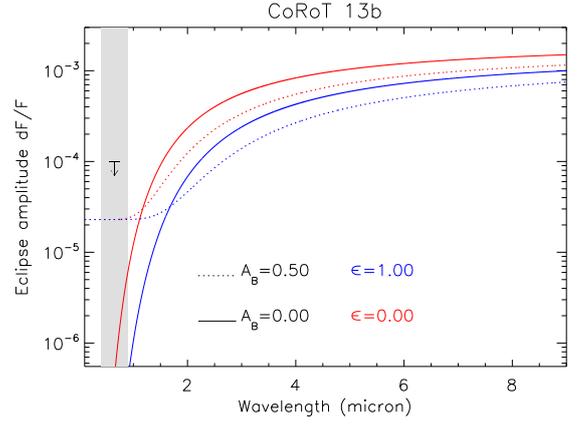}
  \end{center}
  \caption{
% Hans comment 28.5.2010
%%     Expected contrast in the secondary eclipse as a function of the
%%     wavelength for two different values of the Bond albedo and the
%%     heat redistribution factor for CoRoT-13b. The gray shaded area
%%     is the wavelength window observed by \corot. The black arrow is
%%    the upper limit of \corot\,observations.
    Expected amplitude of the occultation of the planet as a function
    of the wavelength for two extreme values for the heat
    redistribution factor for CoRoT-13b (as defined by
    \citealt{cowan2010}; $\epsilon=1$: uniform heat distribution over
    the planet; $\epsilon=0$ no distribution) and two values of the
    Bond albedo. The gray shaded area is the wavelength window
    observed by \corot. The black arrow is the upper limit of
    \corot\,observations. 
  }
  \label{fig:contrast}
\end{figure}

%% % Hans comment 1.6.2010
%% The search for occultations (secondary transits) is challenging in
%% such faint targets. This measurement was possible in CoRoT-1b: the
%% host star has a similar temperature to CoRoT-13, but the star is
%% brighter and the orbital period is only 1.5 days, so the planet is
%% much hotter and the signal-to-noise ratio more favorable
%% \citep{alonso2009a,snellen2009}. 
%% % Hans comment 28.5.2010
%% %% Using for CoRoT-13b the definition for the average surface temperature
%% %% from \citet{cowan2010}, in the most favorable case (zero albedo, zero
%% %% heat  redistribution factor), we obtain a value around 1700K;
%% %% therefore the thermal emission contrast of the occultation should be
%% %% below $4 \cdot 10^{-5}$ (see Fig.~\ref{fig:contrast}). 
%% %% In this temperature range, the emission of the planet in the visible
%% %% range observed by \corot\, is dominated by reflexion.
%% %% Albeit assuming with $A_{B}=0.5$ a very high  Bond albedo, an
%% %% occultation is not expected to be detectable in the current data, as 
%% %% the scatter measured in the folded and binned light curve is of the
%% %% order of $10^{-4}$.
%% %% % (which corresponds to a surface temperature of $2\,500$ K).
%% %% % making such detection impossible using the currently available data.
In the visible wavelength range, occultations of planets behind their
host stars have to date been detected only in a few cases: 
Corot-1b \citep{snellen2009,alonso2009a} and CoRoT-2b
\citep{alonso2009b,snellen2010} in data from \corot\, and HAT-P-7 in 
data from Kepler \citep{borucki2009}.
All are inflated giant planets ($R_{p}>1.4$\Rjup) on very short
orbits, which means that that they have large surfaces favoring a 
large flux of reflected light and additionally they are very hot
($T_{\rm eff} > 2000$K), which favors thermal emission in the
visible regime.
Using the definition from \citet{cowan2010}, CoRoT-13b, in the most
favorable case (zero albedo, zero heat redistribution factor; upper
solid line in Fig.~\ref{fig:contrast}), should have an average surface
temperature around $1700$K.
The emission of the planet in the visible range observed by \corot\,is 
therefore dominated by reflection, meaning that eclipse amplitudes
larger than $2.5\cdot 10^{-5}$ cannot be expected, even with a very
high Bond albedo of $A_{B} = 0.5$ (Fig.~\ref{fig:contrast}, dotted
lines). 
Such a signal is however not detectable in the current data, as the
measured scatter in the folded and binned light curve is on the order
of $10^{-4}$.

%............................................................................
\subsection{Constraints on the presence of additional planets.}
\label{subsec:additional_planets}

\begin{figure}[t]
  \begin{center}
    \includegraphics[%
      width=0.9\linewidth,%
      height=0.5\textheight,%
      viewport=50 50 410 302,%
      keepaspectratio]{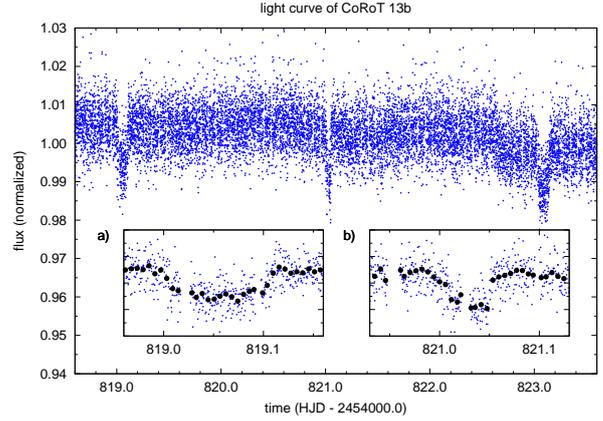}
  \end{center}
  \caption{
    Raw light curve of CoRoT-13b between the observed transits
    number 8 (HJD=$2\,454\,819.06$) and 9 (HJD=$2\,454\,823.09$). 
    a) is a detail of the 8th transit. 
    b) is a detail of the event at HJD=$2\,454\,821.03$ 
    which mimics the transit of a long period planetary companion.
  }
  \label{fig:additionalevent_detail}
\end{figure}

\begin{figure}[t]
  \begin{center}
    \includegraphics[%
      width=0.9\linewidth,%
      height=0.5\textheight,%
%      bb=11 173 600 618,clip,%
      viewport=11 173 600 618,%
      keepaspectratio]{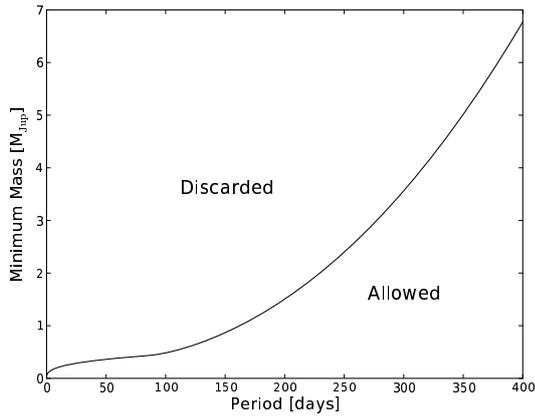}
%      keepaspectratio]{discarded_planets_ecc}
  \end{center}
  \caption{
    Parameter space of the additional planets in the system which can
    been discarded with current HARPS measurements. The region above
    the curve is discarded by RV measurements, while planets located
    in the region below the curve would produce an undetectable
    signal.
  }
  \label{fig:discardedplanets}
\end{figure}

The region of the light curve (LC) shown in
Fig.~\ref{fig:additionalevent_detail}, between the 8th and the 9th
transits observed by \corot\, shows a feature which mimics a single
planetary transit not related to the pattern of transits of
CoRoT-13b. Almost in the middle point of the two 
%faithful 

regular transits we find a short feature, roughly 1.5h long and 1\%
deep which in principle could be due to a planetary transit from an
additional planet in the system. 
% Hans comment 28.5.2010
%% This feature does not repeat in the LC  
%% %(1\% depth is well above the noise level),
%% so we must admit that the hypothetical companion has a period longer
%% than the 80 days that remain until the end of the
%% observations. 
This feature does not repeat in the LC so that a hypothetical
companion would need a period longer than the 80 days that remained
until either end of the observations. 
Nevertheless, the duration of a planetary transit is
linked to the period of the orbit \citep{seager2003} and for an 80
days orbit we expect a transit around 8 hours long. Of course, some
corrections have to be done in the case the orbit is eccentric
\citep{carter2008,kipping2008a}; but in our case we would have to 
admit an eccentricity of 0.94 (and a favorable orientation) for the
hypothetical planet to have a 1.5h long transit in such a long period
orbit. However, this hypothetical planet would have a periastron
passage inside the orbit of CoRoT-13b. There are studies of the
planetary three body problem with high eccentricities (see for example 
\citealt{beauge2003,michtchenko2006}), but to the knowledge of the
authors there is no possible justification for the dynamical stability
of one system formed by a close-in Jupiter-like planet in a circular
orbit and a highly eccentric Jupiter-sized companion.

Moreover, there is no hint of the drift produced by any additional
planet in the scatter of the residuals of the circular radial velocity
(RV) analysis. The region of the parameter space of an additional
planet in a circular orbit which can be discarded based on 
%this criterion alone 
RV data is shown in Fig~\ref{fig:discardedplanets}. We note
that for periods below 85 days (the span of HARPS observations), only
planets in circular orbits with masses below $\sim 0.45$ \Mjup\ are
not discarded by RV data. The mass limit for eccentric orbits is even
lower in this range of periods, as the amplitude $K$ of the RV
movement grows as $(1-e^2)^{-1/2}$.

Given these facts, the most simple explanation is an environmental
effect that mimics a transit. 

\section{Summary}
\label{sec:summary}

We have reported the discovery of CoRoT-13b, a transiting giant
planet orbiting the star \corot\,110839339. 

%% replaced with Malcolm's text
%% We conclude from the spectroscopic analysis that CoRoT-13 is a young 
%% star with solar metallicity and super solar lithium abundance. This
%% fact has been put into the context of the research done by 
%% \citet{israelian2009} and has consequences on the size of the
%% convective zone of the star. 
The spectroscopic analysis reveals that CoRoT-13 is a G0V star with
$T_\mathrm{eff}=5\,945$K, 
$M_{*}=1.09$\Msun, 
$R_{*} = 1.01 \pm 0.03$\Rsun, 
solar metallicity (${\rm [M/H]} = +0.01 \pm 0.07$) 
and a high relative abundance of lithium ($+1.45$ dex). 
The evolutionary tracks constrain the age of the star between $0.12$
and $3.15$ Gyr.  
The results of the study of the lithium abundance, the \vsini\, and
the activity level (which is low both in the spectroscopic and in the
photometric analysis) are consistent with the spectral type and the
age range of the star; although the lower limit of the interval is not
favored due to the low activity level measured. % in the star.

CoRoT-13b is the highest density jovian planet among its relatives
with masses between 0.2 and 2.5 Jupiter masses. Its extreme density
implies the existence of a significant amount, between $140$ and
$300$\Mearth, of heavy elements in the planet, which is contradiction
with the expected link between $M_Z$ of a planet and the metalliciy of
its host stars \citep{guillot2006,burrows2007,guillot2008}. 

There is no hint of any other massive companion in the system, which
is quite a common case for hot Jupiters. 
% We can set an upper limit to the surface temperature of the planet in $2\,500$ K. 

We have discussed the consequences of this characterization and we
show evidence that this particular target provides important
information for the knowledge of the interactions between the
structure of stars, the formation of planets, and their joint
evolution
\citep{ikoma2006,ammler2009,melendez2009a,nordlund2009,ramirez2009,lanza2010a}.

%% Transit surveys are biased towards planets with large radii and hence
%% lower densities.
%% We might be unveiling the high-density part of the planet distribution
%% with \corot\,high-quality light curves. 

%
%____________________________________________________________________________
\begin{acknowledgements}

The team at IAC acknowledges support by grant ESP2007-65480-C02-02 of
the Spanish Ministerio de Ciencia e Innovaci{\'o}n.
This research has made use of the ExoDat database, operated at
LAM-OAMP, Marseille, France, on behalf of the CoRoT/Exoplanet
program. 
This publication makes use of data products from the Two Micron All
Sky Survey, which is a joint project of the University of
Massachusetts and the Infrared Processing and Analysis
Center/California Institute of Technology, funded by the National
Aeronautics and Space Administration and the National Science
Foundation.
This research has made use of NASA's Astrophysics Data System.

%http://www.eso.org/sci/observing/policies/publications.html
%Please notify Uta Grothkopf at esodata@eso.org upon acceptance or
%publication of a paper based on ESO data, including the bibliographic
%reference (article title, authors, journal title, volume, year,
%pages) and the program IDs of the data used. 

\end{acknowledgements}

%
%________________________________________________________________
\bibliographystyle{bibtex/aa}
%\bibliography{/user/ots44/cabr_ju/bibliografia/bibl}
%\bibliography{/home/juano/bibliografia/bibl}
\bibliography{bibl}

\begin{thebibliography}{71}
\expandafter\ifx\csname natexlab\endcsname\relax\def\natexlab#1{#1}\fi

\bibitem[{{Allen}(1973)}]{allen1973}
{Allen}, C.~W. 1973, {Astrophysical quantities} (London: University of London,
  Athlone Press, |c1973, 3rd ed.)

\bibitem[{{Alonso} {et~al.}(2009{\natexlab{a}}){Alonso}, {Alapini}, {Aigrain},
  {Auvergne}, {Baglin}, {Barbieri}, {Barge}, {Bonomo}, {Bord{\'e}}, {Bouchy},
  {Chaintreuil}, {de La Reza}, {Deeg}, {Deleuil}, {Dvorak}, {Erikson},
  {Fridlund}, {de Oliveira Fialho}, {Gondoin}, {Guillot}, {Hatzes}, {Jorda},
  {Lammer}, {L{\'e}ger}, {Llebaria}, {Magain}, {Mazeh}, {Moutou}, {Ollivier},
  {P{\"a}tzold}, {Pont}, {Queloz}, {Rauer}, {Rouan}, {Schneider}, \&
  {Wuchterl}}]{alonso2009a}
{Alonso}, R., {Alapini}, A., {Aigrain}, S., {et~al.} 2009{\natexlab{a}}, \aap,
  506, 353

\bibitem[{{Alonso} {et~al.}(2009{\natexlab{b}}){Alonso}, {Guillot}, {Mazeh},
  {Aigrain}, {Alapini}, {Barge}, {Hatzes}, \& {Pont}}]{alonso2009b}
{Alonso}, R., {Guillot}, T., {Mazeh}, T., {et~al.} 2009{\natexlab{b}}, \aap,
  501, L23

\bibitem[{{Ammler-von Eiff} {et~al.}(2009){Ammler-von Eiff}, {Santos}, {Sousa},
  {Fernandes}, {Guillot}, {Israelian}, {Mayor}, \& {Melo}}]{ammler2009}
{Ammler-von Eiff}, M., {Santos}, N.~C., {Sousa}, S.~G., {et~al.} 2009, \aap,
  507, 523

\bibitem[{{Auvergne} {et~al.}(2009){Auvergne}, {Bodin}, {Boisnard}, {Buey},
  {Chaintreuil}, \& {CoRoT team}}]{auvergne2009}
{Auvergne}, M., {Bodin}, P., {Boisnard}, L., {et~al.} 2009, ArXiv e-prints

\bibitem[{{Baglin} {et~al.}(2006){Baglin}, {Auvergne}, {Boisnard}, {Lam-Trong},
  {Barge}, {Catala}, {Deleuil}, {Michel}, \& {Weiss}}]{baglin2006}
{Baglin}, A., {Auvergne}, M., {Boisnard}, L., {et~al.} 2006, in COSPAR, Plenary
  Meeting, Vol.~36, 36th COSPAR Scientific Assembly, 3749

\bibitem[{{Baliunas} {et~al.}(1995){Baliunas}, {Donahue}, {Soon}, {Horne},
  {Frazer}, {Woodard-Eklund}, {Bradford}, {Rao}, {Wilson}, {Zhang}, {Bennett},
  {Briggs}, {Carroll}, {Duncan}, {Figueroa}, {Lanning}, {Misch}, {Mueller},
  {Noyes}, {Poppe}, {Porter}, {Robinson}, {Russell}, {Shelton}, {Soyumer},
  {Vaughan}, \& {Whitney}}]{baliunas1995}
{Baliunas}, S.~L., {Donahue}, R.~A., {Soon}, W.~H., {et~al.} 1995, \apj, 438,
  269

\bibitem[{{Baraffe} {et~al.}(2008){Baraffe}, {Chabrier}, \&
  {Barman}}]{baraffe2008}
{Baraffe}, I., {Chabrier}, G., \& {Barman}, T. 2008, \aap, 482, 315

\bibitem[{{Baranne} {et~al.}(1996){Baranne}, {Queloz}, {Mayor}, {Adrianzyk},
  {Knispel}, {Kohler}, {Lacroix}, {Meunier}, {Rimbaud}, \& {Vin}}]{baranne1996}
{Baranne}, A., {Queloz}, D., {Mayor}, M., {et~al.} 1996, \aaps, 119, 373

\bibitem[{{Barge} {et~al.}(2008){Barge}, {Baglin}, {Auvergne}, \& {the CoRoT
  team}}]{barge2008b}
{Barge}, P., {Baglin}, A., {Auvergne}, M., \& {the CoRoT team}. 2008, in IAU
  Symposium, Vol. 249, IAU Symposium, 3--16

\bibitem[{{Barnes}(2007)}]{barnes2007}
{Barnes}, S.~A. 2007, \apj, 669, 1167

\bibitem[{{Beaug{\'e}} \& {Michtchenko}(2003)}]{beauge2003}
{Beaug{\'e}}, C. \& {Michtchenko}, T.~A. 2003, \mnras, 341, 760

\bibitem[{{Boisnard} \& {Auvergne}(2006)}]{boisnard2006}
{Boisnard}, L. \& {Auvergne}, M. 2006, in ESA Special Publication, Vol. 1306,
  ESA Special Publication, ed. M.~{Fridlund}, A.~{Baglin}, J.~{Lochard}, \&
  L.~{Conroy}, 19

\bibitem[{{Bord{\'e}} {et~al.}(2010){Bord{\'e}}, {Bouchy}, {Deleuil},
  {Cabrera}, {Jorda}, {Lovis}, {Csizmadia}, {Aigrain}, {Almenara}, {Alonso},
  {Auvergne}, {Baglin}, {Barge}, {Benz}, {Bonomo}, {Bruntt}, {Carone},
  {Carpano}, {Deeg}, {Dvorak}, {Erikson}, {Ferraz-Mello}, {Fridlund},
  {Gandolfi}, {Gillon}, {Guenther}, {Guillot}, {Guterman}, {Hatzes}, {Lammer},
  {L{\'e}ger}, {Mayor}, {Mazeh}, {Moutou}, {P{\"a}tzold}, {Pepe}, {Ollivier},
  {Queloz}, {Rauer}, {Rouan}, {Samuel}, {Santerne}, {Schneider}, {Tingley},
  {Udry}, {Weingrill}, \& {Wuchterl}}]{borde2010}
{Bord{\'e}}, P., {Bouchy}, F., {Deleuil}, M., {et~al.} 2010, \aap

\bibitem[{{Borucki} {et~al.}(2009){Borucki}, {Koch}, {Jenkins}, {Sasselov},
  {Gilliland}, {Batalha}, {Latham}, {Caldwell}, {Basri}, {Brown},
  {Christensen-Dalsgaard}, {Cochran}, {DeVore}, {Dunham}, {Dupree}, {Gautier},
  {Geary}, {Gould}, {Howell}, {Kjeldsen}, {Lissauer}, {Marcy}, {Meibom},
  {Morrison}, \& {Tarter}}]{borucki2009}
{Borucki}, W.~J., {Koch}, D., {Jenkins}, J., {et~al.} 2009, Science, 325, 709

\bibitem[{{Bruntt} {et~al.}(2010{\natexlab{a}}){Bruntt}, {Bedding}, {Quirion},
  {Lo Curto}, {Carrier}, {Smalley}, {Dall}, {Arentoft}, {Bazot}, \&
  {Butler}}]{bruntt2010a}
{Bruntt}, H., {Bedding}, T.~R., {Quirion}, P., {et~al.} 2010{\natexlab{a}},
  ArXiv e-prints

\bibitem[{{Bruntt} {et~al.}(2004){Bruntt}, {Bikmaev}, {Catala}, {Solano},
  {Gillon}, {Magain}, {Van't Veer-Menneret}, {St{\"u}tz}, {Weiss}, {Ballereau},
  {Bouret}, {Charpinet}, {Hua}, {Katz}, {Ligni{\`e}res}, \&
  {Lueftinger}}]{bruntt2004}
{Bruntt}, H., {Bikmaev}, I.~F., {Catala}, C., {et~al.} 2004, \aap, 425, 683

\bibitem[{{Bruntt} {et~al.}(2010{\natexlab{b}}){Bruntt}, {Deleuil}, {Fridlund},
  {Alonso}, {Bouchy}, {Hatzes}, {Mayor}, {Moutou}, \& {Queloz}}]{bruntt2010b}
{Bruntt}, H., {Deleuil}, M., {Fridlund}, M., {et~al.} 2010{\natexlab{b}}, ArXiv
  e-prints

\bibitem[{{Burrows} {et~al.}(2007){Burrows}, {Hubeny}, {Budaj}, \&
  {Hubbard}}]{burrows2007}
{Burrows}, A., {Hubeny}, I., {Budaj}, J., \& {Hubbard}, W.~B. 2007, \apj, 661,
  502

\bibitem[{{Carter} {et~al.}(2008){Carter}, {Yee}, {Eastman}, {Gaudi}, \&
  {Winn}}]{carter2008}
{Carter}, J.~A., {Yee}, J.~C., {Eastman}, J., {Gaudi}, B.~S., \& {Winn}, J.~N.
  2008, \apj, 689, 499

\bibitem[{{Castro} {et~al.}(2009){Castro}, {Vauclair}, {Richard}, \&
  {Santos}}]{castro2009}
{Castro}, M., {Vauclair}, S., {Richard}, O., \& {Santos}, N.~C. 2009, \aap,
  494, 663

\bibitem[{{Chaboyer}(1998)}]{chaboyer1998}
{Chaboyer}, B. 1998, in IAU Symposium, Vol. 185, New Eyes to See Inside the Sun
  and Stars, ed. {F.-L.~Deubner, J.~Christensen-Dalsgaard, \& D.~Kurtz}, 25--+

\bibitem[{{Cowan} \& {Agol}(2010)}]{cowan2010}
{Cowan}, N.~B. \& {Agol}, E. 2010, ArXiv e-prints

\bibitem[{{Deeg} {et~al.}(2009){Deeg}, {Gillon}, {Shporer}, {Rouan},
  {Stecklum}, {Aigrain}, {Alapini}, {Almenara}, {Alonso}, {Barbieri}, {Bouchy},
  {Eisl{\"o}ffel}, {Erikson}, {Fridlund}, {Eigm{\"u}ller}, {Handler}, {Hatzes},
  {Kabath}, {Lendl}, {Mazeh}, {Moutou}, {Queloz}, {Rauer}, {Rabus}, {Tingley},
  \& {Titz}}]{deeg2009}
{Deeg}, H.~J., {Gillon}, M., {Shporer}, A., {et~al.} 2009, \aap, 506, 343

\bibitem[{{Deeg} {et~al.}(2010){Deeg}, {Moutou}, {Erikson}, {Csizmadia},
  {Tingley}, {Barge}, {Bruntt}, {Havel}, {Aigrain}, {Almenara}, {Alonso},
  {Auvergne}, {Baglin}, {Barbieri}, {Benz}, {Bonomo}, {Bord{\'e}}, {Bouchy},
  {Cabrera}, {Carone}, {Carpano}, {Ciardi}, {Deleuil}, {Dvorak},
  {Ferraz-Mello}, {Fridlund}, {Gandolfi}, {Gazzano}, {Gillon}, {Gondoin},
  {Guenther}, {Guillot}, {Hartog}, {Hatzes}, {Hidas}, {H{\'e}brard}, {Jorda},
  {Kabath}, {Lammer}, {L{\'e}ger}, {Lister}, {Llebaria}, {Lovis}, {Mayor},
  {Mazeh}, {Ollivier}, {P{\"a}tzold}, {Pepe}, {Pont}, {Queloz}, {Rabus},
  {Rauer}, {Rouan}, {Samuel}, {Schneider}, {Shporer}, {Stecklum}, {Street},
  {Udry}, {Weingrill}, \& {Wuchterl}}]{deeg2010}
{Deeg}, H.~J., {Moutou}, C., {Erikson}, A., {et~al.} 2010, \nat, 464, 384

\bibitem[{{Deleuil} {et~al.}(2009){Deleuil}, {Meunier}, {Moutou}, {Surace},
  {Deeg}, {Barbieri}, {Debosscher}, {Almenara}, {Agneray}, {Granet},
  {Guterman}, \& {Hodgkin}}]{deleuil2009}
{Deleuil}, M., {Meunier}, J.~C., {Moutou}, C., {et~al.} 2009, \aj, 138, 649

\bibitem[{{D{\'e}sert} {et~al.}(2009){D{\'e}sert}, {Lecavelier des Etangs},
  {H{\'e}brard}, {Sing}, {Ehrenreich}, {Ferlet}, \&
  {Vidal-Madjar}}]{desert2009}
{D{\'e}sert}, J., {Lecavelier des Etangs}, A., {H{\'e}brard}, G., {et~al.}
  2009, \apj, 699, 478

\bibitem[{{Drummond} {et~al.}(2008){Drummond}, {Lapeyrere}, {Auvergne},
  {Vandenbussche}, {Aerts}, {Samadi}, \& {Costa}}]{drummond2008}
{Drummond}, R., {Lapeyrere}, V., {Auvergne}, M., {et~al.} 2008, \aap, 487, 1209

\bibitem[{{Fridlund} {et~al.}(2010){Fridlund}, {H{\'e}brard}, {Alonso},
  {Deleuil}, {Gandolfi}, {Gillon}, {Bruntt}, {Alapini}, {Csizmadia}, {Guillot},
  {Lammer}, {Aigrain}, {Almenara}, {Auvergne}, {Baglin}, {Barge}, {Bord{\'e}},
  {Bouchy}, {Cabrera}, {Carone}, {Carpano}, {Deeg}, {de La Reza}, {Dvorak},
  {Erikson}, {Ferraz-Mello}, {Guenther}, {Gondoin}, {den Hartog}, {Hatzes},
  {Jorda}, {L{\'e}ger}, {Llebaria}, {Magain}, {Mazeh}, {Moutou}, {Ollivier},
  {P{\"a}tzold}, {Queloz}, {Rauer}, {Rouan}, {Samuel}, {Schneider}, {Shporer},
  {Stecklum}, {Tingley}, {Weingrill}, \& {Wuchterl}}]{fridlund2010}
{Fridlund}, M., {H{\'e}brard}, G., {Alonso}, R., {et~al.} 2010, \aap, 512, A14

\bibitem[{{Gandolfi} {et~al.}(2008){Gandolfi}, {Alcal{\'a}}, {Leccia},
  {Frasca}, {Spezzi}, {Covino}, {Testi}, {Marilli}, \&
  {Kainulainen}}]{gandolfi2008}
{Gandolfi}, D., {Alcal{\'a}}, J.~M., {Leccia}, S., {et~al.} 2008, \apj, 687,
  1303

\bibitem[{{Geem} {et~al.}(2001){Geem}, {Kim}, \& {Loganathan}}]{geem2001}
{Geem}, Z.~G., {Kim}, J.~H., \& {Loganathan}, G.~V. 2001, Simulation, 76, 60,
  http://sim.sagepub.com/cgi/content/abstract/76/2/60

\bibitem[{{Gonz{\'a}lez Hern{\'a}ndez} \&
  {Bonifacio}(2009)}]{gonzalezhernandez2009}
{Gonz{\'a}lez Hern{\'a}ndez}, J.~I. \& {Bonifacio}, P. 2009, \aap, 497, 497

\bibitem[{{Guillot}(2005)}]{guillot2005}
{Guillot}, T. 2005, Annual Review of Earth and Planetary Sciences, 33, 493

\bibitem[{{Guillot}(2008)}]{guillot2008}
{Guillot}, T. 2008, Physica Scripta Volume T, 130, 014023

\bibitem[{{Guillot} \& {Morel}(1995)}]{guillot1995}
{Guillot}, T. \& {Morel}, P. 1995, \aaps, 109, 109

\bibitem[{{Guillot} {et~al.}(2006){Guillot}, {Santos}, {Pont}, {Iro}, {Melo},
  \& {Ribas}}]{guillot2006}
{Guillot}, T., {Santos}, N.~C., {Pont}, F., {et~al.} 2006, \aap, 453, L21

\bibitem[{{Ikoma} {et~al.}(2006){Ikoma}, {Guillot}, {Genda}, {Tanigawa}, \&
  {Ida}}]{ikoma2006}
{Ikoma}, M., {Guillot}, T., {Genda}, H., {Tanigawa}, T., \& {Ida}, S. 2006,
  \apj, 650, 1150

\bibitem[{{Israelian} {et~al.}(2009){Israelian}, {Delgado Mena}, {Santos},
  {Sousa}, {Mayor}, {Udry}, {Dom{\'{\i}}nguez Cerde{\~n}a}, {Rebolo}, \&
  {Randich}}]{israelian2009}
{Israelian}, G., {Delgado Mena}, E., {Santos}, N.~C., {et~al.} 2009, \nat, 462,
  189

\bibitem[{{Kipping}(2008)}]{kipping2008a}
{Kipping}, D.~M. 2008, \mnras, 389, 1383

\bibitem[{{Lammer} {et~al.}(2009){Lammer}, {Odert}, {Leitzinger},
  {Khodachenko}, {Panchenko}, {Kulikov}, {Zhang}, {Lichtenegger}, {Erkaev},
  {Wuchterl}, {Micela}, {Penz}, {Biernat}, {Weingrill}, {Steller}, {Ottacher},
  {Hasiba}, \& {Hanslmeier}}]{lammer2009}
{Lammer}, H., {Odert}, P., {Leitzinger}, M., {et~al.} 2009, \aap, 506, 399

\bibitem[{{Lang}(1999)}]{lang1999}
{Lang}, K.~R. 1999, {Astrophysical formulae}, ed. {Lang, K.~R.}

\bibitem[{{Lanza}(2010)}]{lanza2010a}
{Lanza}, A.~F. 2010, \aap, 512, A77

\bibitem[{{Lanza} {et~al.}(2009{\natexlab{a}}){Lanza}, {Aigrain}, {Messina},
  {Leto}, {Pagano}, {Auvergne}, {Baglin}, {Barge}, {Bonomo}, {Collier Cameron},
  {Cutispoto}, {Deleuil}, {De Medeiros}, {Foing}, \& {Moutou}}]{lanza2009b}
{Lanza}, A.~F., {Aigrain}, S., {Messina}, S., {et~al.} 2009{\natexlab{a}},
  ArXiv e-prints

\bibitem[{{Lanza} {et~al.}(2010){Lanza}, {Bonomo}, {Moutou}, {Pagano},
  {Messina}, {Leto}, {Cutispoto}, {Aigrain}, {Alonso}, {Barge}, {Deleuil},
  {Auvergne}, {Baglin}, \& {Collier Cameron}}]{lanza2010b}
{Lanza}, A.~F., {Bonomo}, A.~S., {Moutou}, C., {et~al.} 2010, ArXiv e-prints

\bibitem[{{Lanza} {et~al.}(2009{\natexlab{b}}){Lanza}, {Pagano}, {Leto},
  {Messina}, {Aigrain}, {Alonso}, {Auvergne}, {Baglin}, {Barge}, {Bonomo},
  {Boumier}, {Collier Cameron}, {Comparato}, {Cutispoto}, {de Medeiros},
  {Foing}, {Kaiser}, {Moutou}, {Parihar}, {Silva-Valio}, \&
  {Weiss}}]{lanza2009a}
{Lanza}, A.~F., {Pagano}, I., {Leto}, G., {et~al.} 2009{\natexlab{b}}, \aap,
  493, 193

\bibitem[{{Leconte} {et~al.}(2009){Leconte}, {Baraffe}, {Chabrier}, {Barman},
  \& {Levrard}}]{leconte2009}
{Leconte}, J., {Baraffe}, I., {Chabrier}, G., {Barman}, T., \& {Levrard}, B.
  2009, \aap, 506, 385

\bibitem[{{L{\'e}ger} {et~al.}(2009){L{\'e}ger}, {Rouan}, {Schneider}, {Barge},
  {Fridlund}, {Samuel}, {Ollivier}, {Guenther}, {Deleuil}, {Deeg}, {Auvergne},
  {Alonso}, {Aigrain}, {Alapini}, {Almenara}, {Baglin}, {Barbieri}, {Bruntt},
  {Bord{\'e}}, {Bouchy}, {Cabrera}, {Catala}, {Carone}, {Carpano}, {Csizmadia},
  {Dvorak}, {Erikson}, {Ferraz-Mello}, {Foing}, {Fressin}, {Gandolfi},
  {Gillon}, {Gondoin}, {Grasset}, {Guillot}, {Hatzes}, {H{\'e}brard}, {Jorda},
  {Lammer}, {Llebaria}, {Loeillet}, {Mayor}, {Mazeh}, {Moutou}, {P{\"a}tzold},
  {Pont}, {Queloz}, {Rauer}, {Renner}, {Samadi}, {Shporer}, {Sotin}, {Tingley},
  {Wuchterl}, {Adda}, {Agogu}, {Appourchaux}, {Ballans}, {Baron}, {Beaufort},
  {Bellenger}, {Berlin}, {Bernardi}, {Blouin}, {Baudin}, {Bodin}, {Boisnard},
  {Boit}, {Bonneau}, {Borzeix}, {Briet}, {Buey}, {Butler}, {Cailleau},
  {Cautain}, {Chabaud}, {Chaintreuil}, {Chiavassa}, {Costes}, {Cuna Parrho},
  {de Oliveira Fialho}, {Decaudin}, {Defise}, {Djalal}, {Epstein}, {Exil},
  {Faur{\'e}}, {Fenouillet}, {Gaboriaud}, {Gallic}, {Gamet}, {Gavalda},
  {Grolleau}, {Gruneisen}, {Gueguen}, {Guis}, {Guivarc'h}, {Guterman},
  {Hallouard}, {Hasiba}, {Heuripeau}, {Huntzinger}, {Hustaix}, {Imad},
  {Imbert}, {Johlander}, {Jouret}, {Journoud}, {Karioty}, {Kerjean},
  {Lafaille}, {Lafond}, {Lam-Trong}, {Landiech}, {Lapeyrere}, {Larqu{\'e}},
  {Laudet}, {Lautier}, {Lecann}, {Lefevre}, {Leruyet}, {Levacher}, {Magnan},
  {Mazy}, {Mertens}, {Mesnager}, {Meunier}, {Michel}, {Monjoin}, {Naudet},
  {Nguyen-Kim}, {Orcesi}, {Ottacher}, {Perez}, {Peter}, {Plasson}, {Plesseria},
  {Pontet}, {Pradines}, {Quentin}, {Reynaud}, {Rolland}, {Rollenhagen},
  {Romagnan}, {Russ}, {Schmidt}, {Schwartz}, {Sebbag}, {Sedes}, {Smit},
  {Steller}, {Sunter}, {Surace}, {Tello}, {Tiph{\`e}ne}, {Toulouse}, {Ulmer},
  {Vandermarcq}, {Vergnault}, {Vuillemin}, \& {Zanatta}}]{leger2009}
{L{\'e}ger}, A., {Rouan}, D., {Schneider}, J., {et~al.} 2009, \aap, 506, 287

\bibitem[{{Mamajek} \& {Hillenbrand}(2008)}]{mamajek2008}
{Mamajek}, E.~E. \& {Hillenbrand}, L.~A. 2008, \apj, 687, 1264

\bibitem[{{Mandel} \& {Agol}(2002)}]{mandel2002}
{Mandel}, K. \& {Agol}, E. 2002, \apjl, 580, L171

\bibitem[{{Mayor} {et~al.}(2003){Mayor}, {Pepe}, {Queloz}, {Bouchy},
  {Rupprecht}, {Lo Curto}, {Avila}, {Benz}, {Bertaux}, {Bonfils}, {Dall},
  {Dekker}, {Delabre}, {Eckert}, {Fleury}, {Gilliotte}, {Gojak}, {Guzman},
  {Kohler}, {Lizon}, {Longinotti}, {Lovis}, {Megevand}, {Pasquini}, {Reyes},
  {Sivan}, {Sosnowska}, {Soto}, {Udry}, {van Kesteren}, {Weber}, \&
  {Weilenmann}}]{mayor2003}
{Mayor}, M., {Pepe}, F., {Queloz}, D., {et~al.} 2003, The Messenger, 114, 20

\bibitem[{{Mel{\'e}ndez} {et~al.}(2009{\natexlab{a}}){Mel{\'e}ndez}, {Asplund},
  {Gustafsson}, \& {Yong}}]{melendez2009a}
{Mel{\'e}ndez}, J., {Asplund}, M., {Gustafsson}, B., \& {Yong}, D.
  2009{\natexlab{a}}, \apjl, 704, L66

\bibitem[{{Mel{\'e}ndez} {et~al.}(2009{\natexlab{b}}){Mel{\'e}ndez},
  {Ram{\'{\i}}rez}, {Casagrande}, {Asplund}, {Gustafsson}, {Yong}, {Do
  Nascimento}, {Castro}, \& {Bazot}}]{melendez2009b}
{Mel{\'e}ndez}, J., {Ram{\'{\i}}rez}, I., {Casagrande}, L., {et~al.}
  2009{\natexlab{b}}, \apss, 221

\bibitem[{{Michtchenko} {et~al.}(2006){Michtchenko}, {Beaug{\'e}}, \&
  {Ferraz-Mello}}]{michtchenko2006}
{Michtchenko}, T.~A., {Beaug{\'e}}, C., \& {Ferraz-Mello}, S. 2006, Celestial
  Mechanics and Dynamical Astronomy, 94, 411

\bibitem[{{Mordasini} {et~al.}(2009){Mordasini}, {Alibert}, {Benz}, \&
  {Naef}}]{mordasini2009}
{Mordasini}, C., {Alibert}, Y., {Benz}, W., \& {Naef}, D. 2009, \aap, 501, 1161

\bibitem[{{Morel} \& {Lebreton}(2008)}]{morel2008}
{Morel}, P. \& {Lebreton}, Y. 2008, \apss, 316, 61

\bibitem[{{Nordlund}(2009)}]{nordlund2009}
{Nordlund}, A. 2009, ArXiv e-prints

\bibitem[{{Noyes} {et~al.}(1984){Noyes}, {Hartmann}, {Baliunas}, {Duncan}, \&
  {Vaughan}}]{noyes1984}
{Noyes}, R.~W., {Hartmann}, L.~W., {Baliunas}, S.~L., {Duncan}, D.~K., \&
  {Vaughan}, A.~H. 1984, \apj, 279, 763

\bibitem[{{Pinheiro da Silva} {et~al.}(2008){Pinheiro da Silva}, {Rolland},
  {Lapeyrere}, \& {Auvergne}}]{pinheirodasilva2008}
{Pinheiro da Silva}, L., {Rolland}, G., {Lapeyrere}, V., \& {Auvergne}, M.
  2008, \mnras, 384, 1337

\bibitem[{{Press} {et~al.}(2002){Press}, {Teukolsky}, {Vetterling}, \&
  {Flannery}}]{press2002}
{Press}, W.~H., {Teukolsky}, S.~A., {Vetterling}, W.~T., \& {Flannery}, B.~P.
  2002, Numerical Recipes in C++, 2nd edn. (Cambridge University Press)

\bibitem[{{Queloz} {et~al.}(2009){Queloz}, {Bouchy}, {Moutou}, {Hatzes},
  {H{\'e}brard}, {Alonso}, {Auvergne}, {Baglin}, {Barbieri}, {Barge}, {Benz},
  {Bord{\'e}}, {Deeg}, {Deleuil}, {Dvorak}, {Erikson}, {Ferraz Mello},
  {Fridlund}, {Gandolfi}, {Gillon}, {Guenther}, {Guillot}, {Jorda}, {Hartmann},
  {Lammer}, {L{\'e}ger}, {Llebaria}, {Lovis}, {Magain}, {Mayor}, {Mazeh},
  {Ollivier}, {P{\"a}tzold}, {Pepe}, {Rauer}, {Rouan}, {Schneider},
  {Segransan}, {Udry}, \& {Wuchterl}}]{queloz2009}
{Queloz}, D., {Bouchy}, F., {Moutou}, C., {et~al.} 2009, \aap, 506, 303

\bibitem[{{Ram{\'{\i}}rez} {et~al.}(2009){Ram{\'{\i}}rez}, {Mel{\'e}ndez}, \&
  {Asplund}}]{ramirez2009}
{Ram{\'{\i}}rez}, I., {Mel{\'e}ndez}, J., \& {Asplund}, M. 2009, \aap, 508, L17

\bibitem[{{Sato} {et~al.}(2005){Sato}, {Fischer}, {Henry}, {Laughlin},
  {Butler}, {Marcy}, {Vogt}, {Bodenheimer}, {Ida}, {Toyota}, {Wolf}, {Valenti},
  {Boyd}, {Johnson}, {Wright}, {Ammons}, {Robinson}, {Strader}, {McCarthy},
  {Tah}, \& {Minniti}}]{sato2005}
{Sato}, B., {Fischer}, D.~A., {Henry}, G.~W., {et~al.} 2005, \apj, 633, 465

\bibitem[{{Seager} \& {Mall{\'e}n-Ornelas}(2003)}]{seager2003}
{Seager}, S. \& {Mall{\'e}n-Ornelas}, G. 2003, \apj, 585, 1038

\bibitem[{{Sestito} \& {Randich}(2005)}]{sestito2005}
{Sestito}, P. \& {Randich}, S. 2005, \aap, 442, 615

\bibitem[{{Siess}(2006)}]{siess2006}
{Siess}, L. 2006, \aap, 448, 717

\bibitem[{{Sing}(2010)}]{sing2010}
{Sing}, D.~K. 2010, \aap, 510, A21

\bibitem[{{Snellen} {et~al.}(2009){Snellen}, {de Mooij}, \&
  {Albrecht}}]{snellen2009}
{Snellen}, I.~A.~G., {de Mooij}, E.~J.~W., \& {Albrecht}, S. 2009, \nat, 459,
  543

\bibitem[{{Snellen} {et~al.}(2010){Snellen}, {de Mooij}, \&
  {Burrows}}]{snellen2010}
{Snellen}, I.~A.~G., {de Mooij}, E.~J.~W., \& {Burrows}, A. 2010, \aap, 513,
  A76

\bibitem[{{Surace} {et~al.}(2008){Surace}, {Alonso}, {Barge}, {Cautain},
  {Chabaud}, {Deleuil}, {Fenouillet}, {Meunier}, \& {Moutou}}]{surace2008}
{Surace}, C., {Alonso}, R., {Barge}, P., {et~al.} 2008, in Presented at the
  Society of Photo-Optical Instrumentation Engineers (SPIE) Conference, Vol.
  7019, Society of Photo-Optical Instrumentation Engineers (SPIE) Conference
  Series

\bibitem[{{Torres} {et~al.}(2008){Torres}, {Winn}, \& {Holman}}]{torres2008}
{Torres}, G., {Winn}, J.~N., \& {Holman}, M.~J. 2008, \apj, 677, 1324

\bibitem[{{Winn} {et~al.}(2009){Winn}, {Holman}, {Henry}, {Torres}, {Fischer},
  {Johnson}, {Marcy}, {Shporer}, \& {Mazeh}}]{winn2009}
{Winn}, J.~N., {Holman}, M.~J., {Henry}, G.~W., {et~al.} 2009, \apj, 693, 794

\end{thebibliography}

\end{document}